\documentclass[%
superscriptaddress,
preprint,
 amsmath,amssymb,
 aps,
prb,
]{revtex4-2}

\usepackage{graphicx}% Include Figure~files
\usepackage{dcolumn}% Align table columns on decimal point
\usepackage{bm}% bold math
\usepackage{comment}
\usepackage{units}
\usepackage[T1]{fontenc}
\usepackage{xcolor}
\usepackage{xr}
\usepackage{verbatim}
\usepackage{soul}
\usepackage{microtype}
\usepackage[normalem]{ulem}

\usepackage{placeins}
\usepackage{makecell}
\usepackage{lineno}

\usepackage{hyperref}
\hypersetup{colorlinks=true, allcolors=blue}
\setcitestyle{super,comma,sort&compress}   % superscript, sorted, 1-3 ranges
\makeatletter
\def\@doi#1{\endgroup\href{\doibase#1}{#1}} % clickable DOI without a "doi:" prefix
\makeatother

\begin{document}

\nocite{achemso-control}

\title{The Topography Trap: Sifting Interlayer Excitons from Strain-Related Artifacts in Real-World 2D Heterostructures 
}% Force line breaks with \\

\author{Pablo Hernández López}
\affiliation{Institut für Physik and Center for the Science of Materials Berlin, Humboldt-Universität zu Berlin, Berlin, Germany}
\author{Luka Pirker}
\affiliation{Heyrovsky Institute, Czech Academy of Sciences, Prague, Czech
Republic}
\author{Arijit Kayal}
\affiliation{Heyrovsky Institute, Czech Academy of Sciences, Prague, Czech
Republic}
\author{Astrid Weston}
\affiliation{Department of Physics and Astronomy and National Graphene Institute, University of
Manchester, Manchester, UK}
\author{Rafael Nadas}
\affiliation{Institut für Physik and Center for the Science of Materials Berlin, Humboldt-Universität zu Berlin, Berlin, Germany}
\author{Adrián Dewambrechies Fernández}
\affiliation{Physics Department, Freie Universit\"at Berlin, Berlin, Germany}
\author{Roman Gorbachev}
\affiliation{Department of Physics and Astronomy and National Graphene Institute, University of
Manchester, Manchester, UK}
\author{Kirill I. Bolotin}
\affiliation{Physics Department, Freie Universit\"at Berlin, Berlin, Germany}
\author{Tim Verhagen}
\affiliation{Institute of Physics, Czech Academy of Sciences, Prague, Czech Republic}
\author{Álvaro Rodríguez}
\affiliation{Departamento de Física de la Materia Condensada and Condensed Matter Physics Center (IFIMAC), Universidad Autónoma de Madrid, Madrid, Spain}
\author{Otakar Frank}
\email[Correspondence email address: ]{otakar.frank@heu.cas.cz}
\affiliation{Heyrovsky Institute, Czech Academy of Sciences, Prague, Czech
Republic}
\author{Sebastian Heeg}
\email[Correspondence email address: ]{sebastian.heeg@physik.hu-berlin.de}
\affiliation{Institut für Physik and Center for the Science of Materials Berlin, Humboldt-Universität zu Berlin, Berlin, Germany}

\date{
\today
}% It is always \today, today, 

\maketitle

\newpage
%\linenumbers
\section{Abstract}
Novel excitonic phenomena emerging in transition metal dichalcogenide (TMDC) heterostructures belong to the most exciting topics in contemporary physics of van der Waals materials. Interlayer excitons (IXs) stand out among those due to their long radiative lifetimes and tunability by electric fields, strain, and twist angle. However, many ambiguities persist in the optical identification and manipulation of IXs, highlighting the need for reliable spectroscopic criteria that distinguish interlayer species from spurious signals. Here, we introduce a decision-tree protocol that evaluates interlayer coupling via intralayer exciton quenching and correlates photoluminescence (PL) with atomic force microscopy to establish IX-compatibility of room-temperature PL features in TMDC-based heterostructures. Applied to MoS$_2$--MoSe$_2$ and MoS$_2$--WSe$_2$ heterostructures, it recovers the established momentum-direct IX between the K valleys of the two layers (KK-IX). In WS$_2$--WSe$_2$, a low-energy feature persists through flat, well-coupled regions and for both stacking orders, supporting its assignment as IX-compatible. In contrast, we find no compelling evidence that the widely reported bright, momentum-indirect, twist-angle-independent $\Gamma$K-IX signature in MoS$_2$--WSe$_2$ originates from interlayer emission. Comprehensive experimental data, including infrared and tip-enhanced photoluminescence, stacking order inversion, interface cleaning, control measurements on WSe$_2$/Au, and statistics on variously prepared samples, instead trace these features to locally decoupled and strained WSe$_2$ at interface bubbles and topographical irregularities. Our results expose a generic "topography trap" in real-world heterostructures and establish a practical minimum standard for robust IX assignments, the sift-IX protocol, through spatially resolved room-temperature optical and topographic characterization. 

\section{Keywords}

\noindent Interlayer excitons; TMDC heterostructures; Tip-enhanced photoluminescence; Infrared photoluminescence; Momentum-indirect excitons; Nanobubbles; protocol; sift-IX protocol.

\clearpage

%\begin{nolinenumbers}
\begin{figure}[tp]
	\centering
	\includegraphics[width=0.7\textwidth]{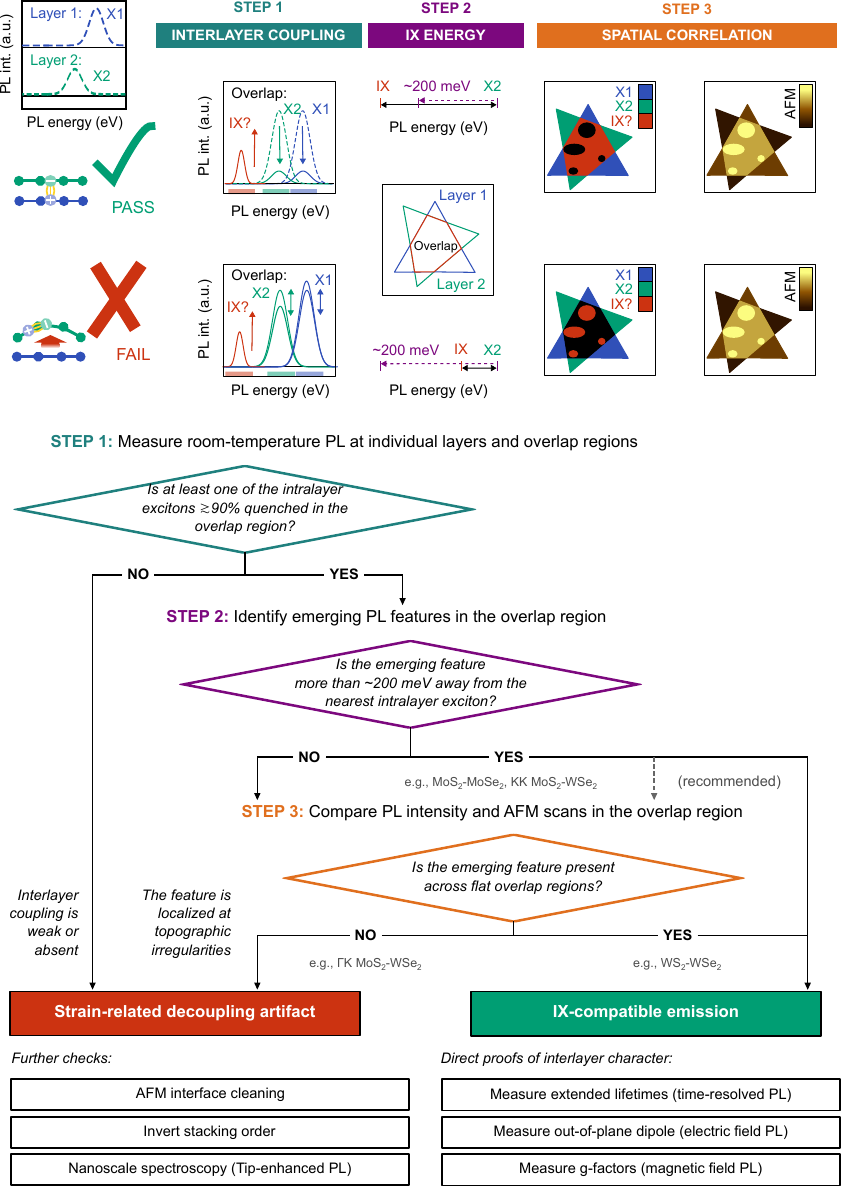}
\caption{\textbf{Screening protocol for interlayer exciton photoluminescence on TMDC-based heterostructures.} Scheme of the sift-IX protocol as a decision tree with the key experimental steps to filter topography-related artifacts and evaluate the compatibility of emergent PL peaks with interlayer emission. We start by collecting single PL spectra at the relevant areas in the sample. Intralayer exciton quenching relative to the isolated monolayers serves as a measure of interlayer coupling. For material combinations where IX lies within an approximate energy range of 200~meV from an intralayer exciton (see Table \ref{table:exampleC}), correlating PL intensity maps and AFM topography allows to determine whether the origin of the emergent PL peaks is compatible with IX emission or a strained intralayer exciton.}
\label{FIG:Protocol}
\end{figure} 
%\end{nolinenumbers}

%\section{Introduction}
Interlayer excitons are bosonic quasiparticles composed of electrons and holes spatially separated in different layers of two-dimensional semiconductor heterostructures. This spatial separation gives rise to a permanent out-of-plane electric dipole moment~\cite{Sigl2022PRBa, Jauregui2019Sa} and extended lifetimes~\cite{Miller2017NLa, Montblanch2021CPa} relative to their intralayer counterparts. IXs are highly tunable in their excitonic characteristics, for example, by selecting the material combination~\cite{Jiang2021LSAa}, strain and twist-angle engineering~\cite{Zhang2024ANa, Nayak2017ANa}, or applying external electric and magnetic fields~\cite{Mak2018NNa,Holler2022PRBa}. Moreover, IXs can be generated, manipulated, and detected optically. These properties have attracted much attention to IXs as a platform for exploring complex correlated many-body states~\cite{Mak2018NNa, Wang2019Na, Tang2020Na, Xu2023PRXa} and for encoding and transmitting information~\cite{Ciarrocchi2018NPa,Rivera2016Sa,Ye2022LS&Aa}.

Optical signatures attributed to IXs have been reported for all combinations of the four main TMDCs, see Table \ref{table:exampleC}, as well as for combinations of these TMDCs and other materials, see Supplementary Note 2. Interlayer excitons emit PL from the infrared to the visible spectral range, and their large binding energies make them stable at room temperature~\cite{Rivera2018NNa}. Depending on whether the electron and hole reside in equivalent or different valleys in their corresponding band structure, IXs can be momentum-direct (e.g., KK) or indirect (e.g., $\Gamma$K or QK), respectively. Momentum-direct IXs are observed only in heterostructures with small twist angles $\theta$, where the K valleys of two layers closely align~\cite{Alexeev2024NLa}. In contrast, momentum-indirect IXs exhibit much weaker twist-angle dependence than their direct counterparts and can be detected at all orientations, relaxing fabrication constraints~\cite{Rivera2018NNa}.

The combination of room-temperature, visible-range photoluminescence (PL) emission, and weak twist-angle dependence (in momentum-indirect IX) makes IXs a highly accessible field across physics, chemistry, and materials science. However, these experimental advantages, together with fabrication-related imperfections such as non-homogeneous strain and chemical residues in dry-stacked, real-world samples~\cite{Covre2022Na, Tilmann2023ANa, Wang2023NEa}, also increase the risks of mischaracterizing emissions resembling IXs. For example, a commonly used polymer, polydimethylsiloxane (PDMS), introduces hydrocarbon-based contaminants at the interface, which coalesce during annealing into bubbles. These often overlooked bubbles give rise to inhomogeneities in the sample topography that may cause misleading PL emission. Conflicting reports exist on the direct or indirect nature of IXs in MoSe$_2$--WSe$_2$~\cite{Rivera2015NCa, Hanbicki2018ANa} and WS$_2$--WSe$_2$ heterostructures~\cite{Jin2019Na,yuan2020twist}, on the hybrid character of WS$_2$--MoSe$_2$ excitons~\cite{Alexeev2019Na, Guo2025NCa} or on the existence of $\Gamma$K interlayer excitons in MoS$_2$--WSe$_2$~\cite{Kunstmann2018NPa, Rodriguez20212Ma} and WS$_2$--WSe$_2$~\cite{Wu2021NSRa, dai2022ScienceAdv}.\\ 

The case of MoS$_2$--WSe$_2$ offers a representative example of the controversies in the field of IX in TMDC heterostructures. Two IXs are reported in MoS$_2$--WSe$_2$: the momentum-direct KK-IX and the momentum-indirect $\Gamma$K-IX. On the one hand, the intravalley KK-IX has been extensively reported in room-temperature and cryogenic PL spectroscopy~\cite{Karni2019PRLa, Liu2019SAa, Chen2024NLa, Tan2022PRLa, Tan2023NMa, Tan2021SAa, Zhao2023NLa} at energies of $\approx$1~eV and directly imaged by ARPES~\cite{Karni2022Na,schmitt2022formation}. Its permanent out-of-plane dipole, direct proof of its interlayer character, has been measured through its Stark shift~\cite{Karni2019PRLa}. On the other hand, the momentum-indirect $\Gamma$K-IX PL in MoS$_2$--WSe$_2$ heterostructures was first reported in 2014~\cite{Fang2014PotNAoSa, Chiu2014ANa}. Because the observed emission energy was above the interlayer bandgap~\cite{Zhang20162Ma, Zhang2017SAa, Ponomarev2018NLa}, it was assigned to a momentum-indirect transition between the K and $\Gamma$ points~\cite{Kunstmann2018NPa, Latini2017NLa}. Many works have followed, using room-temperature PL as the primary evidence for the existence of this novel IX, reported at various energies between 1.5~eV and 1.65~eV~\cite{Unuchek2018Na, Nagler2019pssba, Cho2021NLa, Ren2022APLa, Khestanova2023APa,Imaeda2026TJoPCLa}. Direct proofs of the interlayer character of this IX are, however, missing. Analogously, in WS$_2$--WSe$_2$, a room-temperature $\Gamma$K-IX remains contested~\cite{Wu2021NSRa}, whereas the KK-IX at low temperatures is well established~\cite{Jin2019Na, Jin2019NPa}. These cases exemplify the need for an experimental framework that guides the reliable characterization of IXs in TMDC-based heterostructures. Such guidance must build on widely available, room-temperature methods and instrumentation, as research in new excitonic species between TMDCs and other materials moves forward across multiple disciplines.\\

\begin{table}[h]
\centering
\scriptsize
\renewcommand{\arraystretch}{1.}
\caption{\textbf{Reported energies of intra- and interlayer excitons at room temperature for all combinations of the four main TMDCs and assigned valley transition to the IX.} IX transition assignments with literature discrepancies are highlighted in bold. The energies for the MoSe$_2$ intralayer exciton and IX in WS$_2$--MoSe$_2$ are marked with an asterisk (*) as the observed corresponding excitonic species is reported to be a hybrid of intra- and interlayer exciton.}
\vspace{8pt}
\begin{tabular}{|c|c|c|c|c|c|c|}
\hline
\textbf{\makecell{Combination \\ TMDC1--TMDC2}} &
\textbf{\makecell{Intralayer \\ exciton X1 \\ (eV)}} &
\textbf{\makecell{Intralayer\\ exciton X2 \\ (eV)}} &
\textbf{\makecell{IX\\ (eV)}} &
\textbf{\makecell{Energy\\ separation \\X2-IX (meV)}} &
\textbf{\makecell{IX Transition}} &
\textbf{\makecell{Refs.}} \\
\hline
\textbf{MoS$_2$--MoSe$_2$} & 1.84 & 1.56 & 1.3--1.47 & 90-260 & KK &~\cite{Alexeev2024NLa, rodriguez2023complex,Mouri2017Na} \\
\hline
\textbf{MoS$_2$--WSe$_2$} & 1.86 & 1.64 & 0.9, 1.02 & 650, 620 & KK & ~\cite{Karni2019PRLa,Tan2021SAa, Rodriguez20212Ma} \\
\textbf{} & 1.85 & 1.65 & 1.6 & 50 & $\mathbf{\Gamma}$\textbf{K?} &~\cite{Kunstmann2018NPa, Fang2014PotNAoSa}  \\
\hline
\textbf{WS$_2$--MoS$_2$} & 1.98  & 1.87 & 1.45, 1.52, 1.63 & 420, 350, 240 & $\Gamma$K, QK, KK &~\cite{Gong2014NMa, Okada2018ANa, Schottle2025APa}  \\
\hline
\textbf{WS$_2$--MoSe$_2$} & 1.98 & 1.49--1.57* & 1.49--1.57* & - & \textbf{*Hyb MoSe$_2$-IX} &~\cite{Alexeev2019Na, Zhang2020NCa}\\
\hline
\textbf{WS$_2$--WSe$_2$} & 1.95 & 1.63 & 1.3--1.55 & 80--330 & \makecell{ $\mathbf{\Gamma}$\textbf{K/KK, KK/QK}} &~\cite{Jin2019NPa, Wu2021NSRa, Chen2023RAa}\\
\hline
\textbf{WSe$_2$--MoSe$_2$} & 1.66 & 1.57 & 1.33-1.38 & 190-240 & \textbf{KK/QK} &\cite{Rivera2015NCa,Nayak2017ANa, Tran2019Na, Hanbicki2018ANa,Miller2017NLa} \\
\hline
\end{tabular}
\label{table:exampleC}
\end{table}

In this work, we introduce the sift-IX protocol, an accessible, room-temperature protocol for screening candidate IX-PL features in TMDC heterobilayers (Figure~\ref{FIG:Protocol}). This protocol evaluates compatibility with interlayer emission and filters common strain- and topography-related artifacts, without replacing direct measurements of interlayer character. We illustrate its possible outcomes using several representative systems. First, we recover the established, spectrally well-separated KK-IX emission in MoS$_2$--WSe$_2$ and MoS$_2$--MoSe$_2$. We then examine candidate momentum-indirect IX emission in WS$_2$--WSe$_2$ and MoS$_2$--WSe$_2$, where the spectral proximity to the WSe$_2$ intralayer exciton creates a substantial risk of misassignment. The WS$_2$–WSe$_2$ feature is present throughout flat, well-coupled regions and for both stacking orders, supporting its compatibility with interlayer emission. By contrast, the prominently reported $\Gamma$K-IX-like emission in MoS$_2$--WSe$_2$ is confined to topographical irregularities and also fails the stacking-order check. Tip-enhanced PL with 20~nm spatial resolution, interface cleaning, WSe$_2$ control measurements and statistics across variously prepared MoS$_2$--WSe$_2$ samples collectively trace these suspect PL features to locally decoupled and strained WSe$_2$. Overall, the combination of re-evaluation and protocol presented here is a practical example of addressing the demand for reproducibility in twisted TMDC heterobilayers research~\cite{boggild2025protocols}. 

\section{Results and discussion}%I added these just to make it more readable 
\subsection{The sift-IX protocol for interlayer exciton photoluminescence}

Figure~\ref{FIG:Protocol} presents the sift-IX protocol, our proposed protocol to evaluate the compatibility of room-temperature PL features from TMDC heterostructures with IX emission.  It uses single PL spectra of monolayers and overlap regions,  together with PL mapping in most cases, and AFM topography where required. The aim of the protocol is to guide the characterization of TMDC heterostructures, providing checks to flag artifacts derived from topography rather than substitute direct proofs of the interlayer character of an IX. Moreover, we intend the decision-tree-based workflow of the sift-IX protocol to be highly accessible. The protocol relies on standard experimental techniques such as PL microspectroscopy and atomic force microscopy (AFM), see Methods in Supplementary~Note~1. Its applicability extends beyond combinations of the four main TMDCs and can be adapted to, among others, heterobilayers formed with non-conventional TMDCs (e.g., rhenium-based compounds and tellurides) as well as perovskites and molecules. We include an extended discussion of the applicability of the sift-IX protocol in Supplementary~Note~2. Concerning the four main TMDCs, our criteria hold across the full range of dry-stacking conditions: encapsulated or bare samples, CVD-grown or exfoliated flakes, and any choice of transfer polymer, see Supplementary~Note~3. The main requisite for a heterostructure to host IXs is a type-II band alignment between the constituent monolayers. A type-II or staggered band alignment drives charge transfer between the layers, spatially separating electrons and holes and thereby enabling the formation of IXs~\cite{Rivera2018NNa}. Prior knowledge of the band alignment, as well as the band-structure features of the constituent monolayers, allows us to estimate the expected emission energy range of IXs and thus provides an initial level of screening between genuine IX-like emission and strain-induced PL from individual monolayers.\\

We start the characterization of a heterobilayer in step one of Figure~\ref{FIG:Protocol} by defining three significant areas on the sample: the two isolated monolayers and the overlap area forming the heterostructure. During fabrication, we intentionally avoid full overlap to use the PL of the isolated layers as a reference. In the overlap region, complete or almost complete quenching of intralayer excitons relative to their intensities in the individual layers indicates efficient charge transfer and therefore strong interlayer coupling. We discuss quantitative quenching factors for different material combinations in Supplementary~Note~2. In contrast, the absence of quenching of intralayer excitons represents strong evidence of poor interlayer coupling, questioning any further assignments to interlayer species.\\

After establishing that quenched intralayer emission confirms interlayer coupling, we proceed in Figure~\ref{FIG:Protocol} with step two of the sift-IX protocol, which requires looking for additional excitonic PL signatures in the energy range of the transition corresponding to the expected IX. Here, we aim to distinguish real IX-compatible features from artifacts that may be emitting in the same energy range. Specifically, topographical irregularities in the overlap area can cause locally unquenched intralayer exciton emission that may be mistaken for interlayer species. These locally unquenched intralayer excitons are often downshifted in energy due to strain-driven band-gap reduction~\cite{Niehues2018NLa} and can overlap spectrally with IXs. We propose that the risk of strained intralayer excitons and IX overlapping is particularly high when the unstrained intralayer exciton and IX lie typically within 200~meV for the four common TMDCs under the conditions considered here, requiring additional analysis. Outside the conventional strain-accessible window, topographic correlation is still recommended, but the risk of confusion with strained intralayer emission is significantly reduced. This threshold is not fundamental but reflects typical tensile strains attainable at bubbles or wrinkles (1--2$\%$~\cite{Khestanova2016NCa}) and the corresponding exciton shift rates reported for TMDC monolayers, ranging from 50--140~meV~\cite{Rodriguez20212Ma, Darlington2020TJoCPa, Gastaldo2023n2MaAa}.\\

In step three of the sift-IX protocol, Figure~\ref{FIG:Protocol}, we perform hyperspectral PL mapping and AFM imaging and compare PL intensity maps and AFM topography. Our goal is to examine the spatial origin of all PL emissions within or close to 200~meV from the intralayer excitons. AFM topography reveals topographical features in the overlap area, including trapped contaminant pockets at the interface of the bilayer (or outside of it in the case of encapsulated samples), as well as cracks and folds, which range in size from tens to thousands of nanometers. We create PL maps of the emergent signal and the intralayer excitons, integrating PL intensity in the corresponding energy ranges. If the PL intensity of the emergent signal is only present at the topography irregularities and absent in flat areas, and, at the same time, the intralayer PL quenching is reduced at the location of the candidate emission, we conclude that the emission in question is very unlikely to originate from interlayer species.\\

The spatial resolution of hyperspectral mapping should match the size of the topographical features in the sample to produce unequivocal maps. If topographical features are smaller than the light diffraction limit, spectroscopic techniques capable of reaching the corresponding lateral resolution, like tip-enhanced photoluminescence (TEPL), are necessary. This and other further checks of the spectroscopic evidence of IXs, including stacking order inversion and contact-AFM interface cleaning, are described below and in Supplementary~Note~4, respectively. For PL features that pass all three sifting steps in Figure~\ref{FIG:Protocol}, the sift-IX protocol confirms their compatibility with IX emission and establishes a minimal standard for IX assignments based on room-temperature PL and AFM. This, however, does not constitute a direct, absolute proof of their interlayer character. Orthogonal measurements---including an out-of-plane Stark response~\cite{Jauregui2019Sa}, time-resolved dynamics~\cite{Rivera2015NCa}, and magneto-optical $g$-factors~\cite{Seyler2019Na}---can establish interlayer character with substantially greater confidence.\\

Having established the main analysis steps of the sift-IX protocol in Figure~\ref{FIG:Protocol}, we apply it to three representative systems, MoS$_2$--MoSe$_2$, MoS$_2$--WSe$_2$ and WS$_2$--WSe$_2$. In the following, we will separate the materials in a heterostructure with a dash (--) to indicate arbitrary stacking order and with a slash (/) to indicate that the first material is on top.\\

\subsection{{Interlayer excitons: KK-IX in MoS$_2$--MoSe$_2$ and MoS$_2$--WSe$_2$}}

%\begin{nolinenumbers}
\begin{figure}[tp]
	\centering\includegraphics[width=\textwidth]{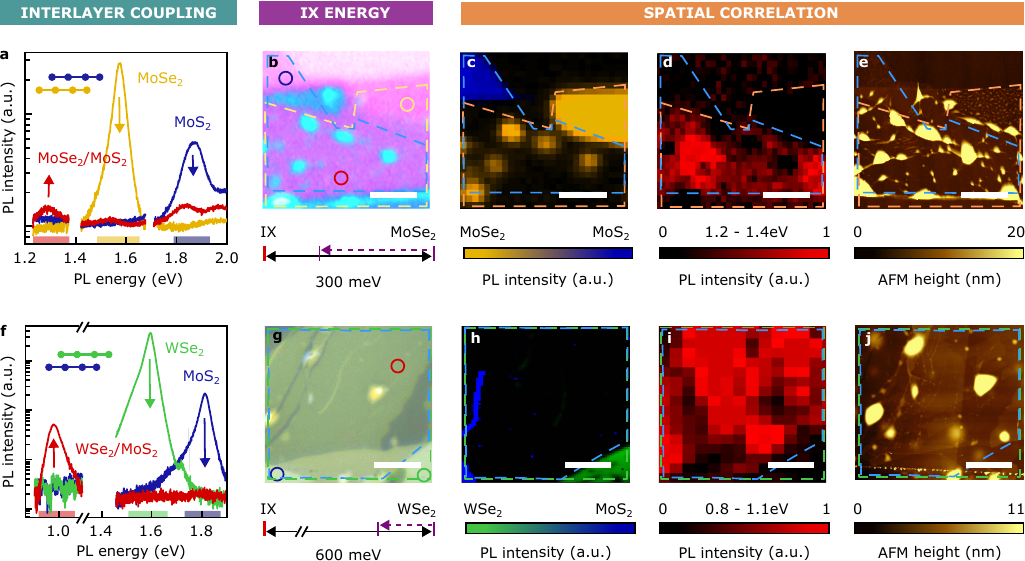}
\caption{\textbf{Screening interlayer excitons in TMDC-based heterostructures: Pass cases} Sequential characterization, following the sift-IX protocol in Figure~\ref{FIG:Protocol}, of KK-IX in MoS$_2$--MoSe$_2$ ({\bf a--e}), and KK-IX in MoS$_2$--WSe$_2$ ({\bf f--j}), both passing all tests. ({\bf a, f})~Representative infrared and visible PL spectra collected at the points marked with the same color in the microscope images of the heterostructures. PL intensity on a log scale. ({\bf b, g}) Microscopic images of the samples and a schematic of the energy distance between IX and the nearest intralayer exciton. Purple arrows mark the $\sim$200~meV threshold proposed in step two of the protocol.  ({\bf c, h})~PL intensity maps of the corresponding intralayer excitons. ({\bf d, i})~PL intensity maps of KK-IX. ({\bf e, j})~AFM topography maps with cropped height range. Scale bars: b,c,d,e: 4~\textmu m, g,h,i,j: 10~\textmu m.}
\label{FIG:KKPass}
\end{figure} 
%\end{nolinenumbers}

In MoS$_2$--MoSe$_2$, an IX between the K valleys of MoS$_2$ and MoSe$_2$ is expected at 1.3~eV for twist angles close to 0$^\circ$ or 60$^\circ$, where K valleys align, and the transition between them is momentum-direct~\cite{rodriguez2023complex}. We present in Figure~\ref{FIG:KKPass}a representative PL spectra from the relevant regions in the aligned MoSe$_2$/MoS$_2$ sample shown in Figure~\ref{FIG:KKPass}b, featuring the corresponding intralayer excitons at 1.9~eV (MoS$_2$, blue) and 1.6~eV (MoSe$_2$, orange) in the isolated monolayers, and a peak at 1.3~eV in the overlap region (red). Intralayer excitons in the overlap area are quenched by 86\% (MoS$_2$) and 99\% (MoSe$_2$) compared to the monolayer regions. This meets the condition in step one of the protocol. A KK-IX is expected at 1.3~eV, 300~meV away from the nearest intralayer emission, the MoSe$_2$ intralayer exciton. This satisfies the condition in step two. Therefore, following Figure~\ref{FIG:Protocol}, the feature is compatible with an IX at 1.3~eV. Indeed, direct proofs of the interlayer character of the KK-IX in MoS$_2$--MoSe$_2$ have been presented in Refs.~\cite{Alexeev2024NLa, Mouri2017Na}. We include for completeness the spatial PL maps of the MoS$_2$ and MoSe$_2$ intralayer excitons, Figure~\ref{FIG:KKPass}c, and the KK-IX, Figure~\ref{FIG:KKPass}d.  KK-IX is present in most of the overlap area but not in the isolated monolayers, also passing step three of the protocol. Moreover, the regions where the KK-IX intensity drops to zero coincide with the only regions in the overlap area where intralayer emission increases, and with interfacial bubbles seen in the micrograph and the AFM scan.\\

Similarly to MoS$_2$--MoSe$_2$, in WSe$_2$--MoS$_2$, an IX between the corners of the Brillouin zones is expected for samples with twist angles close to 0$^\circ$ or 60$^\circ$. Figure~\ref{FIG:KKPass}f shows representative infrared and visible room-temperature PL spectra of an aligned WSe$_2$/MoS$_2$ sample from the points indicated in Figure~\ref{FIG:KKPass}g. Its twist angle is measured to be 6.5~$\pm$~0.1$^\circ$ via second-harmonic generation (SHG) in Supplementary Note 5.

At the individual monolayers, we observe MoS$_2$ (blue, 1.85~eV) and WSe$_2$ (green, 1.66~eV) intralayer excitons. In the overlap area, these intralayer excitons are quenched by >99\% compared to the monolayers, and a new peak emerges at 1~eV, the energy of the KK transition, 600~meV away from the WSe$_2$ intralayer exciton. Following the sift-IX protocol, this emerging peak is fully compatible with IX emission. Its interlayer character has been unequivocally demonstrated by Karni \textit{et al.}~\cite{Karni2019PRLa}. In Supplementary Notes 4 and 5, we further check the robustness of the MoS$_2$--WSe$_2$ KK-IX against stacking order inversion and twist angle variation.  In the PL intensity map in Figure~\ref{FIG:KKPass}h, intralayer emission is present in the individual monolayers but is completely quenched in the heterostructure. In contrast, in Figure~\ref{FIG:KKPass}i, KK-IX emission is present only in the heterostructure region and absent in the isolated monolayers. These observations clearly confirm that the conditions for the formation of IX are fully met everywhere in the heterostructure area. However, we do not observe any PL emission in the spectral region between 1.4~eV and 1.6~eV, where the momentum-indirect $\Gamma$K-IX has been reported~\cite{Kunstmann2018NPa}. We look explicitly for $\Gamma$K-IX, which do not fulfill the 200~meV-threshold, in the following sections.\\

%\begin{nolinenumbers}
\begin{figure}[tp]
	\centering\includegraphics[width=\textwidth]{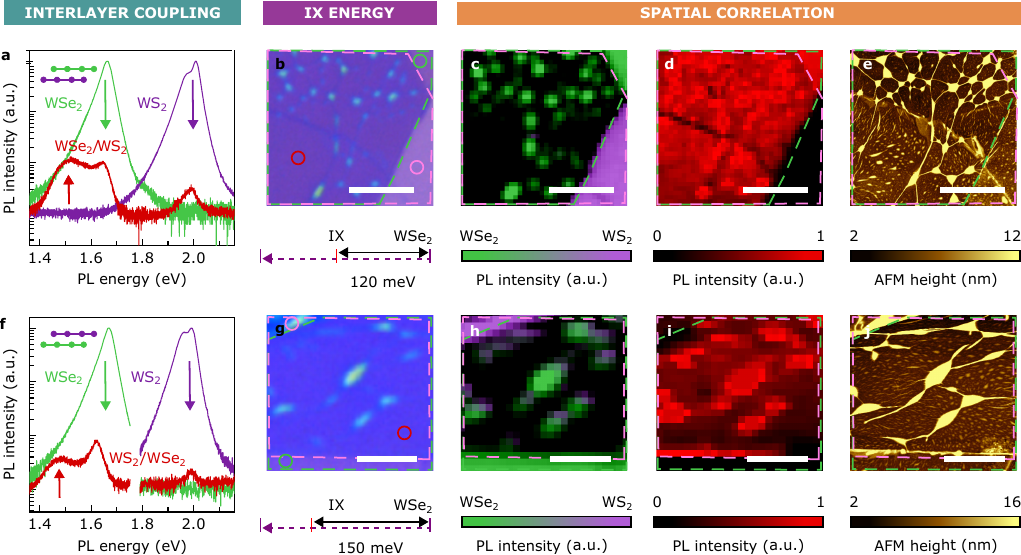}
\caption{\textbf{Screening candidate $\Gamma$K-IX emission in WS$_2$-WSe$_2$: pass case.} Sequential characterization, following the sift-IX  protocol in Figure~\ref{FIG:Protocol}, of $\Gamma$K-IX in WS$_2$--WSe$_2$ with the two possible stacking orders, WSe$_2$ on top, {\bf a--e}, and WS$_2$ on top,  {\bf f--j}. Present in flat areas of both stacking orders, the low-energy feature passes the protocol and is compatible with interlayer emission. ({\bf a, f})~Representative infrared and visible PL spectra collected at the points marked with the same color in the microscope images of the heterostructures. PL intensity on a log scale. ({\bf b, g}) Microscopic images of the samples and a schematic of the energy distance between IX and the nearest intralayer exciton. Purple arrows mark the $\sim$200~meV threshold proposed in step two of the protocol.  ({\bf c, h})~PL intensity maps of the corresponding intralayer excitons. ({\bf d, i})~PL intensity maps of the IX candidate. ({\bf e, j})~AFM topography maps with cropped height range. Scale bars: b,c,d,e: 3~\textmu m, g,h,i,j: 5~\textmu m.}
\label{FIG:WS2WSe2}
\end{figure} 
%\end{nolinenumbers}

%\begin{nolinenumbers}
\begin{figure}[tp]
	\centering\includegraphics[width=\textwidth]{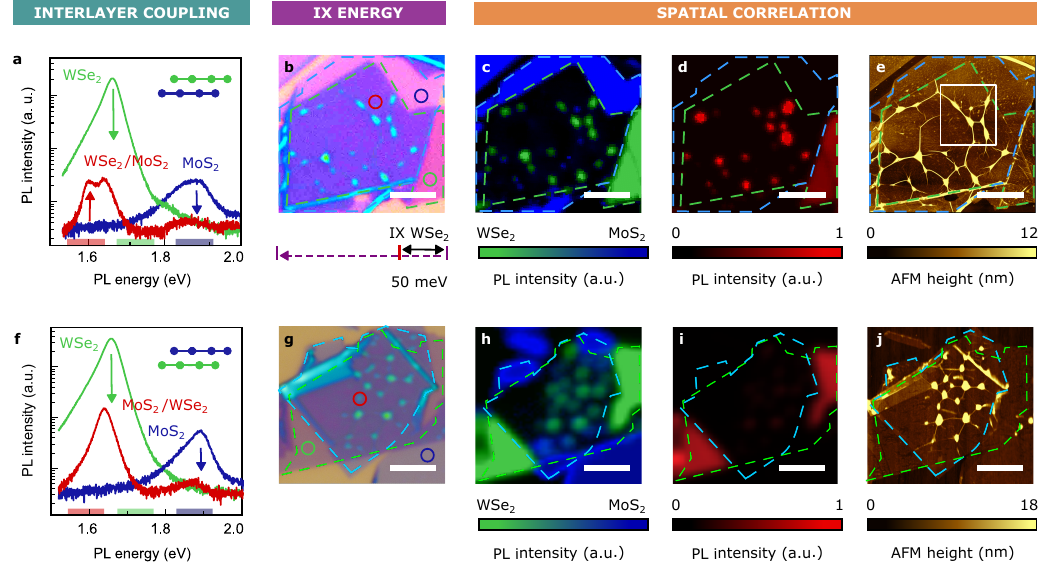}
\caption{\textbf{Screening interlayer excitons in TMDC-based heterostructures: Fail case} The protocol introduced in Figure~\ref{FIG:Protocol} flags the feature commonly attributed to a $\Gamma$K-IX in MoS$_2$--WSe$_2$ as inconsistent with interlayer emission. The emission also fails to pass the additional robustness check against stacking order inversion. ({\bf a, f})~Representative log-scale PL spectra from the points marked in ({\bf b, g}). Microscope images of a WSe$_2$/MoS$_2$ heterostructure (b) and a MoS$_2$/WSe$_2$ heterostructure (g), and a schematic of the energy distance between IX and the nearest intralayer exciton. Purple arrows mark the $\sim$200~meV threshold proposed in step two of the protocol. ({\bf c, h})~PL intensity maps of the corresponding intralayer excitons. ({\bf d, i})~PL intensity maps integrated in the range 1.55--1.62~eV where a $\Gamma$K-IX is usually reported. ({\bf e, j})~AFM topography map with cropped height range. A white square in (e) marks the area analyzed in Figure~\ref{FIG:TEPL}. Scale bars: 3~\textmu m.}
\label{FIG:Kunstmann}
\end{figure} 
%\end{nolinenumbers}

\subsection{Interlayer emission compatible \boldmath{$\Gamma$K-IX} in WS$_2$--WSe$_2$}

We now examine the scenario where the proximity of the candidate IX to the nearest intralayer exciton requires a careful comparison of its spatial distribution and the sample's topography. This corresponds to step three of the sift-IX protocol in Figure~\ref{FIG:Protocol}. We illustrate this step using the momentum-indirect $\Gamma$K-IX. $\Gamma$K-IX has been reported both in WS$_2$--WSe$_2$~\cite{Wu2021NSRa} and MoS$_2$--WSe$_2$~\cite{Kunstmann2018NPa} between an electron in the K valley and a hole in a hybridized $\Gamma$ valley. Unlike KK-IX, $\Gamma$K-IX do not require a specific lattice alignment. Their optical signatures are expected to be present at all twist-angles, at energies that vary with $\theta$. We start by looking for the $\Gamma$K-IX in WS$_2$--WSe$_2$.\\

We fabricate two WS$_2$--WSe$_2$ heterostructures with opposite stacking orders: WSe$_2$ on top (Figure~\ref{FIG:WS2WSe2}a--e) and WS$_2$ on top (Figure~\ref{FIG:WS2WSe2}f--j). Figure~\ref{FIG:WS2WSe2}a,f shows representative spectra from the isolated monolayers and overlap areas collected at the points indicated in Figure~\ref{FIG:WS2WSe2}b,g. We compare the intralayer excitons at the monolayers (WSe$_2$ at 1.66~eV, green, and WS$_2$ at 2~eV, purple) with those in the overlap area (red) and find $>$99\% quenching in all cases. An emerging feature assigned to $\Gamma$K-IX appears as a shoulder 120 and 150~meV below the WSe$_2$ intralayer exciton in the two samples, respectively. Both values lie within the 200~meV threshold suggested in step two of the sift-IX protocol. We therefore continue with the correlation of the PL intensity maps for the intralayer excitons (Figure~\ref{FIG:WS2WSe2}c,h), the emerging feature (Figure~\ref{FIG:WS2WSe2}d,i) and the AFM topography (Figure~\ref{FIG:WS2WSe2}e,j). The PL intensity integration ranges are indicated in Figure~\ref{FIG:WS2WSe2}a,f. Intralayer excitons in Figure~\ref{FIG:WS2WSe2}c,h dominate the emission of the isolated monolayers and the bubbles in the overlap area, but vanish in the overlap regions between bubbles. In contrast, the signal below the WSe$_2$ intralayer exciton is present in the whole overlap area, including the regions between bubbles, Figure~\ref{FIG:WS2WSe2}d,i, and disappears in the WS$_2$ monolayer regions. The PL non-zero intensity integrated in this range, observed at the WSe$_2$ monolayer and the bubbles in the overlap region, leaks from the tail of the WSe$_2$ intralayer exciton. Critically, the emergent feature is present in flat overlap areas, passing the criterion in step three of the protocol, and for both stacking orders, passing also one of the further checks suggested at the end of Figure~\ref{FIG:Protocol}. We therefore conclude that the reported $\Gamma$K-IX in WS$_2$--WSe$_2$ is compatible with interlayer emission and not a topography artifact.

\subsection{Topography artifacts: \boldmath{$\Gamma$K-IX} in MoS$_2$--WSe$_2$}

Finally, we return to MoS$_2$--WSe$_2$ to examine the features that have previously been assigned to $\Gamma$K-IX, applying the sift-IX protocol of Figure~\ref{FIG:Protocol} to two MoS$_2$--WSe$_2$ heterostructures with opposite stacking orders.

Figure~\ref{FIG:Kunstmann}a shows representative PL spectra from the locations indicated in Figure~\ref{FIG:Kunstmann}b. Compared to the monolayer regions, the spectrum from the overlap area (red) features 99\% (WSe$_2$) and 85\% (MoS$_2$) quenched intralayer excitons PL and an additional PL peak at 1.59~eV, which may be interpreted as the indirect $\Gamma$K-IX. The emerging feature lies only 50~meV below the WSe$_2$ intralayer exciton. Following step three in the protocol, we compare hyperspectral maps (Figures~\ref{FIG:Kunstmann}c,d) and AFM topography (Figure~\ref{FIG:Kunstmann}e) to accurately determine the origin of the peak. WSe$_2$ and MoS$_2$ intralayer excitons dominate the individual layers and are almost fully quenched in a large part of the heterostructure region. However, the PL peak emerging around 1.59~eV (red spots in Figure~\ref{FIG:Kunstmann}d) appears only when the intralayer PL remains unquenched (green spots in Figure~\ref{FIG:Kunstmann}c) and is absent otherwise.\\

Indeed, there is a clear spatial correlation between the unquenched WSe$_2$ intralayer PL, Figure~\ref{FIG:Kunstmann}c, the PL feature at 1.59~eV, Figure~\ref{FIG:Kunstmann}d, and the bubbles in the AFM topography,  Figure~\ref{FIG:Kunstmann}e. The same pattern is also apparent in the microscope image, Figure~\ref{FIG:Kunstmann}b. Bubbles locally decouple WSe$_2$ from the underlying MoS$_2$, reducing PL quenching, and induce strain, which redshifts the intralayer emission from the top layer. In Supplementary Note 6, we demonstrate that the same scenario leads to similar PL features in WSe$_2$ bubbles on a gold substrate.

At this point, the sift-IX  protocol flags the PL feature at 1.59~eV as inconsistent with interlayer emission. We now move on to implement the additional checks suggested by the protocol. We invert the stacking order of the sample in Figure~\ref{FIG:Kunstmann}b and fabricate a MoS$_2$/WSe$_2$ heterostructure, Figure~\ref{FIG:Kunstmann}g, with representative PL spectra shown in Figure~\ref{FIG:Kunstmann}f. This time, in contrast to $\Gamma$K-IX in WS$_2$-WSe$_2$ in the previous section, after inverting the stacking order, we find no emission below the WSe$_2$ intralayer exciton that can be assigned to an IX. We compare the PL hyperspectral maps (Figure~\ref{FIG:Kunstmann}h,i) with the AFM topography (Figure~\ref{FIG:Kunstmann}j) and observe enhanced intralayer emission in regions coincident with bubbles visible both in the micrograph and AFM scan, but no emission in the flat areas. Moreover, there is almost no emission in the 1.55--1.62~eV range even at those bubbles with bright intralayer excitons, in contrast to the WSe$_2$/MoS$_2$ sample, Figure~\ref{FIG:Kunstmann}d, where 1.55--1.62~eV PL intensity increases at the bubbles. Interfacial bubbles strain mainly the top layer. When WSe$_2$ is on top, its intralayer emission is therefore redshifted into the energy window expected for the IX, whereas when WSe$_2$ is at the bottom, its emission barely shifts across the interface. We observe this across many samples and expand the argument with sample statistics below. This stacking order check indicates that the features assigned to the $\Gamma$K-IX in MoS$_2$--WSe$_2$ are, in fact, strain-related artifacts.

\subsection{TEPL of a WSe$_2$/MoS$_2$ heterostructure}

%\begin{nolinenumbers}
\begin{figure}[ht]
	\centering
	\includegraphics[width=8.6cm]{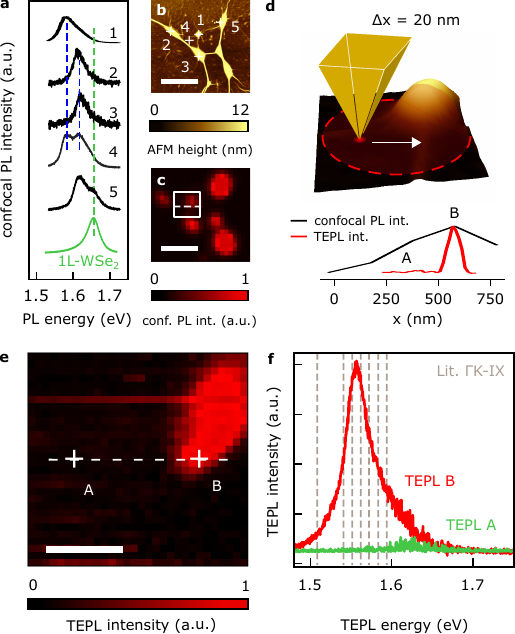}
    \caption{\textbf{Resolving the topography trap with TEPL.} TEPL provides the nanoscale spatial resolution required when the sift-IX protocol encounters sub-diffraction-limited topographical features, linking the emergent low-energy PL to strained WSe$_2$ rather than to an IX. ({\bf a})~Representative confocal PL spectra collected at positions marked in (b) and unstrained WSe$_2$ for reference (green). Blue dashed lines: curve 4 consists of overlapping contributions from bubbles measured in curves 1, 2, and 3. ({\bf b})~Detail of AFM topography of the sample in Figure~\ref{FIG:Kunstmann}e. ({\bf c})~Confocal PL intensity map between 1.45–-1.65 eV for the same region. ({\bf d})~Schematic of the TEPL measurement. The area shaded in red corresponds to the confocal laser spot. Inset: line cuts of confocal and TEPL intensity across a bubble corresponding to dashed lines in (c) and (e), respectively. ({\bf e})~TEPL map at the white square in (c). ({\bf f})~Representative TEPL spectra from (e). Grey dashed lines: Reported energies misassigned to a $\Gamma$K-IX. From left to right: Refs.~\cite{Cho2021NLa, Fang2014PotNAoSa, Khestanova2023APa, Unuchek2018Na, Ren2022APLa, Kunstmann2018NPa, Chiu2014ANa, Cho2021NLa}. Scale bars are 1.5~\textmu m (b,c) and 100~nm (d).}
    %.~[59, 33, 61, 57, 60, 27, 52] 

\label{FIG:TEPL}
\end{figure} 
%\end{nolinenumbers}

We now focus on individual spectra that may be misinterpreted as emerging PL features different from the intralayer WSe$_2$ exciton due to a lack of resolution in confocal spectroscopy. Figure~\ref{FIG:TEPL}a shows exemplary micro--PL spectra from the bubbles marked in Figure~\ref{FIG:TEPL}b. Spectra near bubbles may feature one shifted peak, as in curves 1, 2, or 3, or two distinct peaks, as in curves 4 and 5. This double-peak feature occurs when the laser spot probes areas with varying strain, such as two differently strained bubbles, thereby collecting emission from both. For example, blue dashed lines in Figure~\ref{FIG:TEPL}a mark how curve 4 includes overlapping contributions from the neighboring bubbles measured in curves 1, 2, and 3. The observed downshifts of 20--80~meV correspond to 0.2--0.7\% biaxial tensile strain~\cite{kumar2024strain} and shift the WSe$_2$ intralayer exciton energy into the range at which the indirect IX has previously been reported. Note that a comparable analysis of the double peak is not available in the $\Gamma$K-IX literature, which largely neglects a thorough spatial characterization that compares PL maps and microscope images or AFM topography.~\cite{Liu2019SAa, Chen2024NLa, Tan2022PRLa, Rosenberger2020ANa, Tan2023NMa, Tan2021SAa, Zhao2023NLa}.\\

To overcome the limitations of confocal spectroscopy, we illustrate an additional spectroscopic check proposed at the end of the sift-IX protocol in Figure~\ref{FIG:Protocol}, performing TEPL with nanoscale spatial resolution on one of the contaminant bubbles visible in Figure~\ref{FIG:Kunstmann}b-e. TEPL unequivocally localizes the PL signal commonly attributed to $\Gamma$K-IX exclusively to a WSe$_2$ bubble. We look at the $\sim$200~nm bubble labeled 1 in Figure~\ref{FIG:TEPL}b. This bubble produces the lowest energy PL feature of the sample, see curve 1 in Figure~\ref{FIG:TEPL}a. We perform TEPL in the region marked with a white square in Figure~\ref{FIG:TEPL}c. The gain in lateral resolution is apparent when comparing a $20$~nm-resolved TEPL line scan across the bubble with the corresponding confocal PL and AFM topography line scans, Figure~\ref{FIG:TEPL}d. The TEPL intensity map is shown in Figure~\ref{FIG:TEPL}e. Representative TEPL spectra along the line profile next to (A) and on (B) the bubble are shown in Figure~\ref{FIG:TEPL}f. Combined, they show that redshifted PL originates exclusively from the strained WSe$_2$ bubble, while PL from the surrounding, well-coupled areas is completely quenched. This scenario also holds true for the opposite stacking order, where MoS$_2$ emission on top of the bubbles is redshifted and WSe$_2$ emission remains at the same energy, while in the surrounding flat areas emission is completely quenched, as we show in Supplementary~Note~4.\\*

%\begin{nolinenumbers}
\begin{figure}[tp]
	\centering
	\includegraphics[width=0.98\textwidth]{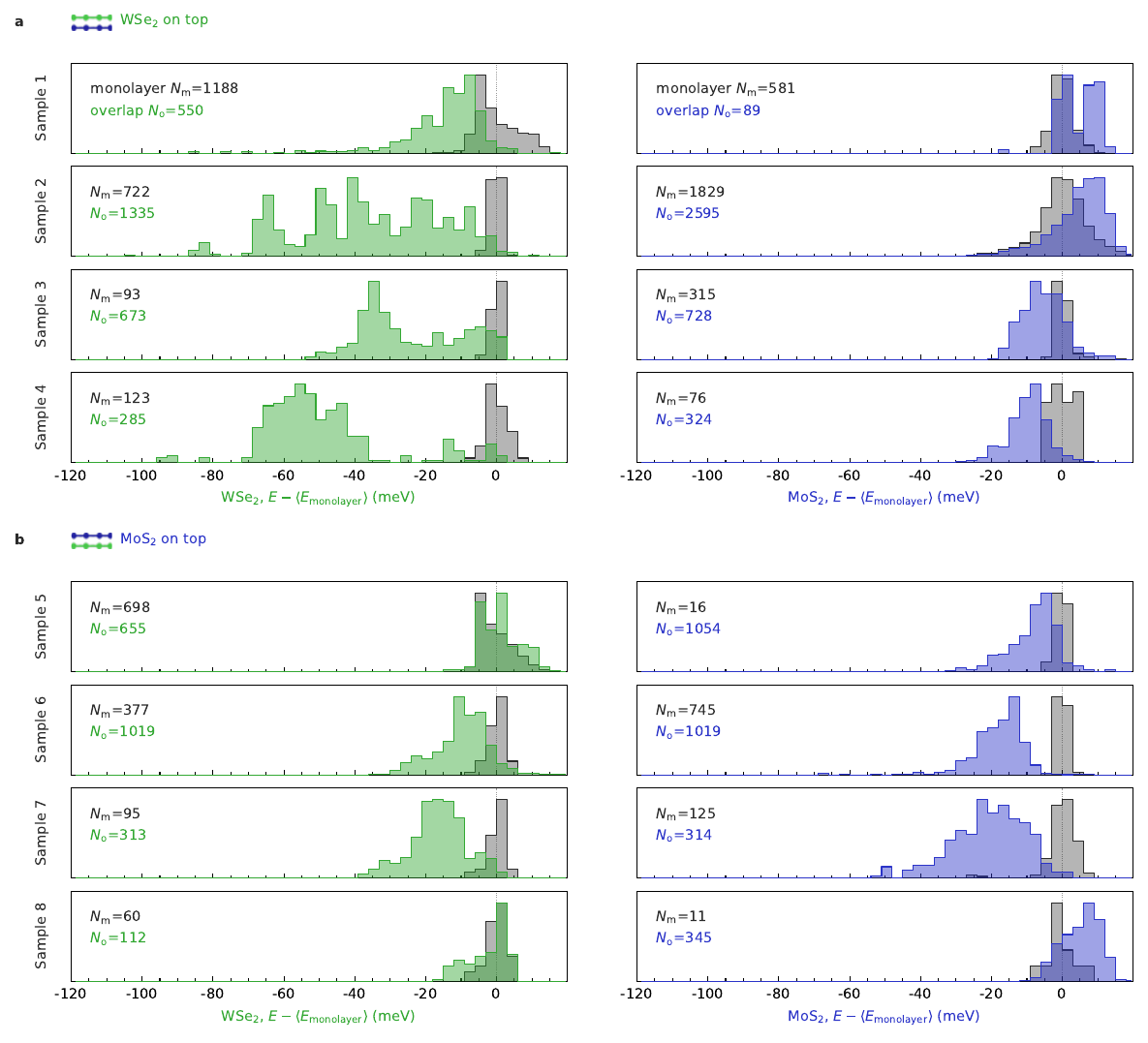}
\caption{\textbf{PL energy distribution of MoS$_2$--WSe$_2$ heterostructures} PL energy distribution histograms in the WSe$_2$ (left, green) and MoS$_2$ (right, blue) range in the monolayer and overlap regions for eight MoS$_2$--WSe$_2$ heterostructures sorted by stacking order: WSe2 on top \textbf{(a)} and MoS$_2$ on top \textbf{(b)}. Energy is plotted as the distance to the average energy in the monolayer region, in meV. The number of spectra considered is indicated in the top left corner.}
\label{FIG:Stats}
\end{figure} 
%\end{nolinenumbers}

Local decoupling and straining of WSe$_2$ at topographical irregularities produce a big diversity of PL signals in energy and shape. Crucially, as it is shown with the dashed vertical lines in Figure~\ref{FIG:TEPL}f, the energy range of these strain-induced PL features almost perfectly matches the energy range reported for the absent $\Gamma$K-IX~in the literature. We analyze in Figure~\ref{FIG:Stats} the energy distribution in the monolayer and overlap regions of eight MoS$_2$--WSe$_2$ samples divided by stacking order. As a reference, monolayer regions (grey) show narrow emission energy dispersions, around 5--10\,meV. For WSe$_2$, dispersion and redshift of the emission energy are two and five times bigger, respectively, when WSe$_2$ is on top (Figure~\ref{FIG:Stats}a) than when MoS$_2$ is on top (Figure~\ref{FIG:Stats}b), with energies spanning a range down to 100~meV below the unstrained intralayer exciton. The details of our statistical analysis can be found in Supplementary~Note~3. On the one hand, the energy of PL from strained WSe$_2$ may depend indirectly on sample parameters like substrate adhesion, which influences the bubble strain~\cite{Khestanova2016NCa}, as we also show in Supplementary~Note~6, where WSe$_2$ monolayers on gold exhibit redshifts of up to 220~meV, or twist angle, which modifies the interlayer coupling via atomic reconstruction~\cite{Weston2020NNa}. On the other hand, the PL spectral shape from strained WSe$_2$ may vary with strain inhomogeneity, as it was analyzed above for curve 4 in Figure~\ref{FIG:TEPL}a, which depends on bubble size, shape, and distribution, as well as on other effects like strain-induced hybridization~\cite{HernandezLopez2022NCa} and trion conversion~\cite{harats2020dynamics}.\\*

\subsection{Discussion}

Applying the sift-IX protocol to the features hitherto identified as $\Gamma$K-IX in MoS$_2$--WSe$_2$ heterobilayers reveals a clear incompatibility with interlayer emission. A summary and extended discussion of our arguments against assigning the 1.5--1.65 eV feature in MoS$_2$--WSe$_2$ to a bright $\Gamma$K-IX is provided in Supplementary~Note~7. Given the absence of compelling evidence that the emission in this range originates from a bright, momentum-indirect, twist-angle-independent IX in this system, our study calls for efforts to clarify the nature of experimental observations attributed to this species. This pertains to alleged properties such as tunability by mechanical strain~\cite{Cho2021NLa}, dominating radiative decay after ultrafast photo-excited charge transfer~\cite{Zimmermann2021ANa}, robustness against charge-carrier doping~\cite{Khestanova2023APa}, as well as the foundational demonstration of a room-temperature excitonic transistor based on the electric control of $\Gamma$K-IX flux~\cite{Unuchek2018Na}.\\*

Beyond MoS$_2$–WSe$_2$, our work establishes a practical, room-temperature framework for screening common artifacts related to sample topography in spectroscopy of TMDC heterostructures. In its minimal form, a robust IX assignment should satisfy three criteria: (i) substantial quenching of intralayer excitons (typically $\gtrsim 90\%$ in at least one layer) demonstrating efficient interlayer charge transfer~\cite{Rivera2015NCa,You2022APLa}, (ii) an IX energy that lies outside the strain-accessible redshift window of intralayer excitons (or, if within $\sim 200$~meV, is rigorously disentangled from strain-induced features), and (iii) an emergent PL signal that does not spatially correlate with topography-induced decoupling such as bubbles, wrinkles, or cracks. When these criteria are met, additional experiments that directly probe the interlayer character—such as lifetime measurements, Stark shifts, or characteristic $g$-factors—can elevate an IX-compatible assignment to a confirmed IX~\cite{Karni2019PRLa,Sigl2022PRBa,Seyler2019Na,Jauregui2019Sa,Miller2017NLa}.\\*

The detailed spatial characterizations suggested in our sift-IX protocol, combining hyperspectral PL mapping, AFM topography, and, where necessary, TEPL with nanometer resolution, are also instrumental in advancing our understanding of confirmed IX within heterogeneous strain and dielectric landscapes. They provide direct insight into how local strain, interfacial cleanliness, and reconstruction reshape the excitonic potential landscape, which is particularly relevant in moiré superlattices and other engineered quantum materials~\cite{Alexeev2019Na,Karni2022Na,Tan2022PRLa,Rosenberger2020ANa}. As device concepts increasingly rely on controlled exciton funnels, moiré trapping, and long-range IX transport, such spatially resolved protocols become a prerequisite rather than an optional refinement.\\*

By providing an accessible and standardized methodology based on widely available optical and topographic techniques, we aim to enhance the robustness and reproducibility of results derived from real-world heterostructures. The "topography trap" highlighted here is unlikely to be unique to MoS$_2$–WSe$_2$; it is a generic risk whenever interlayer species are inferred solely from PL at room temperature in samples with imperfect interfaces. The sift-IX protocol therefore offers a generally applicable framework for screening candidate interlayer-exciton emission in 2D semiconductor heterostructures and for preventing artifacts from seeding entire research lines in this vibrant and rapidly expanding field.

\section{Conclusions}

We have introduced the sift-IX protocol, an accessible, room-temperature decision-tree protocol that combines intralayer-exciton quenching, energy placement relative to a realistic strain window, and spatial correlation of PL with AFM topography to sift IX-compatible emission from strain- and topography-induced artifacts in TMDC-based heterostructures. Applied to MoS$_2$–MoSe$_2$, and MoS$_2$–WSe$_2$ bilayers, the sift-IX protocol robustly identifies the established momentum-direct KK-IX emission. In WS$_2$–WSe$_2$, we validate 1.4--1.5~eV emission as also interlayer-compatible, but find no compelling evidence for a bright, momentum-indirect, twist-angle–independent $\Gamma$K-IX in MoS$_2$–WSe$_2$. Instead, the corresponding PL features originate from locally decoupled and strained WSe$_2$ at interface bubbles and other topographical irregularities, exemplifying a generic “topography trap” in real-world 2D heterostructures.

Beyond this specific case, our work establishes a minimal standard for IX assignments based on room-temperature PL and AFM. This framework is readily applicable to other TMDC and moiré heterostructures, where local strain, reconstruction, and dielectric disorder reshape the excitonic landscape, and it complements direct probes of interlayer character. As the field moves toward scalable devices and complex excitonic architectures, adopting such protocol-driven standards will be essential to ensure that reported interlayer species are robust, reproducible, and suitable as reliable building blocks for future quantum and optoelectronic technologies.

\section*{Supporting Information}
Experimental methods; Applicability of the sift-IX protocol; Sample statistics of the topography trap; IX screening methods beyond the sift-IX protocol; Twist-angle-dependence test of MoS$_2$--WSe$_2$ and WS$_2$--WSe$_2$ heterostructures; Interface bubbles between WSe$_2$ and a gold substrate; Summary of arguments against the reported $\Gamma$K-IX assignment in MoS$_2$--WSe$_2$.

\section{Author contributions}
P.H.L., O.F., and S.H. conceived the study. P.H.L., L.P., A.K. and A.R. fabricated the samples. A.W. fabricated the encapsulated samples with support from R.G. P.H.L., L.P. and A.K. performed the PL and AFM measurements. A.D.F. and K.B. assisted in preliminary PL characterization. P.H.L., L.P. and R.N. performed the TEPL measurements. L.P. and T.V. performed the SHG measurements. P.H.L. analysed the data together with L.P. and A.K. P.H.L., O.F., and S.H. co-wrote the manuscript with input from all authors. O.F. and S.H. supervised the project. All authors have given approval to the final version of the manuscript.

\section{Acknowledgments}
We acknowledge support by the German Research Foundation (DFG) and the Open Access Publication Fund of Humboldt-Universität zu Berlin. A.K. and O.F. acknowledge the Czech Science Foundation project (GAČR 23-06174K). L.P. and O.F. acknowledge the support of the Ministry of Education, Youth, and Sports of the Czech Republic, Project No. CZ.02.01.01/00/22\_008/0004558, co-funded by the European Union. T.V. acknowledges the support provided by the Ferroic Multifunctionalities project, supported by the Ministry of Education, Youth, and Sports of the Czech Republic, Project No. CZ.02.01.01/00/22\_008/0004591, co-funded by the European Union. P.H.L., R.N., and S.H. acknowledge funding from the Deutsche Forschungsgemeinschaft (DFG, German Research Foundation) under the Emmy Noether Initiative (Project-ID 433878606). R.N. and S.H. acknowledge the DFG within the CRC 1772 mol2Dmat project (project numbers B02; project ID 555467911). K.B. acknowledges funding from BMFTR (05K2022 ioARPES) and DFG (CRC 1772, project B01, SPP2244).\\
The authors used Perplexity (Perplexity AI) to improve the grammar and readability of the manuscript; all suggested edits were reviewed and revised by the authors. Claude (Anthropic) was used to assist in writing Python scripts for processing and plotting data, and to generate the schematic sieve in the TOC graphic. The authors verified the code and its outputs against manual measurements and take full responsibility for the content of this publication.

\section{Data availability}
All data supporting the key findings of this study shown in the article and the Supplementary Information file are available in Zenodo at https://doi.org/10.5281/zenodo.23732895. All raw data generated during the current study are available from the corresponding author upon request. Source data are provided with this paper.

\clearpage
% Don't count these!
%TC:ignore
%\quickwordcount{main_resubmission}
%\quickcharcount{MoS2metals}
%\detailtexcount{main_resubmission}

\clearpage

\bibliographystyle{achemso}
\bibliography{acs-control,20261020TopographyTrap_fixed}

%\clearpage
%\section*{For Table of Contents Only}
%\begin{center}
%\includegraphics[width=3.25in,height=1.75in,keepaspectratio]{./Figures/20261027_TOC_v6.pdf}
%\end{center}

\end{document}

\nocite{achemso-control}

\title{Supporting Information:\\The Topography Trap: Sifting Interlayer Excitons from Strain-Related Artifacts in Real-World 2D Heterostructures}

% Authors in indicated order (not grouped)

\author{Pablo Hernández López}
\affiliation{Institut für Physik and Center for the Science of Materials Berlin, Humboldt-Universität zu Berlin, Berlin, Germany}
\author{Luka Pirker}
\affiliation{Heyrovsky Institute, Czech Academy of Sciences, Prague, Czech
Republic}
\author{Arijit Kayal}
\affiliation{Heyrovsky Institute, Czech Academy of Sciences, Prague, Czech
Republic}
\author{Astrid Weston}
\affiliation{Department of Physics and Astronomy and National Graphene Institute, University of
Manchester, Manchester, UK}
\author{Rafael Nadas}
\affiliation{Institut für Physik and Center for the Science of Materials Berlin, Humboldt-Universität zu Berlin, Berlin, Germany}
\author{Adrián Dewambrechies Fernández}
\affiliation{Physics Department, Freie Universit\"at Berlin, Berlin, Germany}
\author{Roman Gorbachev}
\affiliation{Department of Physics and Astronomy and National Graphene Institute, University of
Manchester, Manchester, UK}
\author{Kirill I. Bolotin}
\affiliation{Physics Department, Freie Universit\"at Berlin, Berlin, Germany}
\author{Tim Verhagen}
\affiliation{Institute of Physics, Czech Academy of Sciences, Prague, Czech Republic}
\author{Álvaro Rodríguez}
\affiliation{Departamento de Física de la Materia Condensada and Condensed Matter Physics Center (IFIMAC), Universidad Autónoma de Madrid, Madrid, Spain}
\author{Otakar Frank}
\email[Correspondence email address: ]{otakar.frank@heu.cas.cz}
\affiliation{Heyrovsky Institute, Czech Academy of Sciences, Prague, Czech
Republic}
\author{Sebastian Heeg}
\email[Correspondence email address: ]{sebastian.heeg@physik.hu-berlin.de}
\affiliation{Institut für Physik and Center for the Science of Materials Berlin, Humboldt-Universität zu Berlin, Berlin, Germany}

\date{\today}

 \maketitle

\tableofcontents
\listoffigures
\listoftables
\clearpage
\onecolumngrid
\linenumbers

%\section{About the Protocol}
\section{Supplementary Note~1: Experimental Methods}

\subsection{Sample fabrication} 
Two different fabrication strategies were employed for the heterobilayers, determining the final sample quality. Both strategies lead to similar results, as we show statistically in Supplementary Note~3. Standard samples, intended for comparison with existing literature, were fabricated via a widely accessible PDMS (polydimethylsiloxane)-mediated dry transfer technique. This group includes the MoSe$_2$--MoS$_2$ and WS$_2$--WSe$_2$ heterostructures and the majority of MoS$_2$--WSe$_2$, see Supplementary Note~3. For these samples, monolayers were mechanically exfoliated using Nitto tape (BT-130E-SL) and transferred onto commercial PDMS (Gel-Pak X4) stamps. Stacks were then assembled on Si/SiO$_2$ substrates (or a 170~\textmu m thick glass coverslip for the sample in Supplementary Figure \ref{MoS2TEPL}, see Optical measurements below) using a custom-built transfer system equipped with micromanipulators and a resistive heating unit. In the case of the MoS$_2$--MoSe$_2$ heterostructure, the twist angle was selected by aligning the straight crystallographic edges of both layers. The presence of IX confirms $0^{\circ}$/$60^{\circ}$ twist alignment. For selected samples indicated in Supplementary Note~5, the twist angle was additionally measured using second-harmonic generation (SHG). The heterobilayers were subsequently annealed either on a hotplate at 200$^{\circ}$C for 3 hours or at 150$^{\circ}$C for 1 hour inside a glovebox to promote the aggregation of interlayer contaminants in bubbles, leaving pristine areas behind. Both annealing methods yielded similar results.\\*

For high-quality samples with large-scale pristine areas and precise crystallographic alignment (near $0^{\circ}$/$60^{\circ}$), a PC (polycarbonate)-coated PDMS pick-up method was utilized. Monolayers of MoS$_2$ and WSe$_2$ were exfoliated from bulk crystals using Nitto tape on a hotplate at 60$^{\circ}$C, then transferred onto PVA (polyvinyl alcohol)/ PMMA (polymethyl methacrylate)-coated Si/SiO$_2$ wafers. Thick hBN flakes ($\approx$60~nm) were exfoliated directly on Si/SiO$_2$. The flakes were sequentially picked up using a PC stamp within a nitrogen-filled glovebox, utilizing a motorized transfer system for precise angle control. The twist angle between layers was selected by aligning the straight edges of the flakes, confirmed by the presence of KK-IX, and measured via SHG in the sample in Figure~2. The final heterostructure was released onto a Si/SiO$_2$ substrate, and the PC was washed away with chloroform at room temperature. The samples were then annealed in vacuum at 200$^{\circ}$C for an hour. Finally, residual bubbles were pushed aside when possible using a SiN cantilever, squeezing bubbles out at an angle of 20-40$^{\circ}$~\cite{Wang2023NEa}.\\*

\subsection{AFM characterization and interface cleaning}

AFM characterization of the samples' topography was performed in a Horiba OmegaScope AFM in tapping mode with ACCESS-SNC-G probes ($k$ = 4.45~N/m, $f_0$ = 125~kHz) from APP-NANO and in a Dimension ICON AFM in Peak Force Tapping mode with SCANASYST-AIR AFM probes ($k$ = 0.4~N/m, $f_0$ = 70~kHz), both from Bruker.

The contact-mode AFM interface cleaning described in Supplementary Note~4 was implemented in a JPK Nanowizzard 2 AFM system following the procedure first introduced in Ref.~\cite{Rosenberger2018AAM&Ia}. To clear out a certain area from contaminants, contact-mode probes (BudgetSensors ContAl-G, $k$ = 0.2~N/m, $f_0$ = 13~kHz) were used to apply forces between 60--90~nN in several successive scans (up to nine in Figure~\ref{Squeegee}). Samples before and after the treatment were imaged in contact mode using an applied force of 2~nN.

%\begin{figure}[tp]
%	\centering
%	\includegraphics[width=0.98\textwidth]{./Figures/FigureS6.pdf}
%\caption[Full-scale AFM topography maps]{\textbf{Full-scale AFM topography maps.} In Figures 2 and 3 of the main text and Figure S2, we show AFM topography maps of the corresponding samples with optimized contrast to highlight the heterostructures' flat interface. Here, we show the complete, uncropped AFM height ranges corresponding to \textbf{a} Figure~2e, \textbf{b} Figure~2j, \textbf{c} Figure~3e, \textbf{d, e} Figures~S5c, f. Scale bars: 4, 10, 3, 2, 2~\textmu m.}
%\label{AFMfull}
%\end{figure} 

\subsection{Optical measurements}

Confocal PL measurements in the visible range (i.e. all confocal PL measurements in Figures~2d, i, n and their corresponding curves in Figures~2a, f, k, Figures 3 and 4, and ~\ref{StackKK}c, f) were performed using commercial systems WITec 360 from Oxford Instruments and XploRA PLUS from Horiba, both with a 532~nm continuous wavelength laser as excitation source and a 100$\times$ (NA = 0.9) objective, 40~\textmu W laser power and 600~gr/mm grating. Confocal infrared PL data were collected using a Labram HR from Horiba equipped with an InGaAs detector (Symphony InGaAs-1700, Horiba), a 50$\times$ (NA = 0.55) objective, and a 633~nm excitation laser.\\*

Tip-enhanced PL measurements in Figure~\ref{MoS2TEPL} were acquired using FabNS Porto system~\cite{rabelo2019tip}, consisting of a micro-Raman setup in an inverted microscope configuration combined with a shear-force scanning probe microscopy (SPM) setup equipped with a gold plasmon-tunable tip pyramid (PTTP) for local enhancement of the optical electric fields~\cite{miranda2020optical}. An 60$\times$ oil-immersion objective (NA = 1.4), a HeNe excitation laser (633~nm) and a 600~grooves (gr)/mm grating were used with laser power of 20~$\mu$W and 1~s acquisition time. The oil immersion objective and inverted microscope configuration require samples to be assembled on 170~\textmu m thick glass coverslips, as described above in Sample fabrication.\\*

TEPL measurements in Figure 3 of the main text were conducted using an AIST-NT AFM system integrated with LabRAM HR Evolution Raman setup from Horiba. TEPL data was recorded with commercially available Ag-coated probes (Omni-TERS-SNC-Ag) from APP-NANO in dual scan mode, where spectra for both tip-sample in-contact and out-of-contact were recorded for each pixel. A 633 nm laser was used as excitation, focused at the tip edge with a 100$\times$ (NA = 0.7) objective. Typical laser power ~150 \textmu W and 150~gr/mm grating was used for the TEPL map with 1~s acquisition time for each spectrum.\\*

Second-harmonic generation (SHG) measurements were performed using a WITec alpha300R Raman microscope with a UHTS600 spectrometer (300 gr/mm), equipped with a 1064~nm pulsed laser (NPI Rainbow 1064 OEM, pulse width $<10$~ps), focused through a Zeiss EC Epiplan $50\times$ objective (NA = 0.75). The laser power at the objective was 3.5~mW. The polarizer and analyzer were oriented perpendicular to each other.
The crystallographic orientation of each constituent monolayer was determined by recording their SHG intensity as a function of the excitation polarization angle $\theta$, and fitting it to $I(\theta) = I_0 \cos^2[3(\theta - \theta_0)]$, where $\theta_0$ corresponds to the lattice's armchair direction. The twist angle of each heterostructure, up to a $0^\circ/60^\circ$ ambiguity, was then obtained from the difference between the armchair directions of its two constituent monolayers. \\*

\subsection{Data treatment}

Data were processed and plotted using custom Python scripts written with the assistance of Claude (Anthropic). A data viewer is available alongside the data in the repository (see Data availability in the main text).\\*

\section{Supplementary Note~2: Applicability of the sift-IX protocol}  

\begin{figure}[tp]
	\centering
	\includegraphics{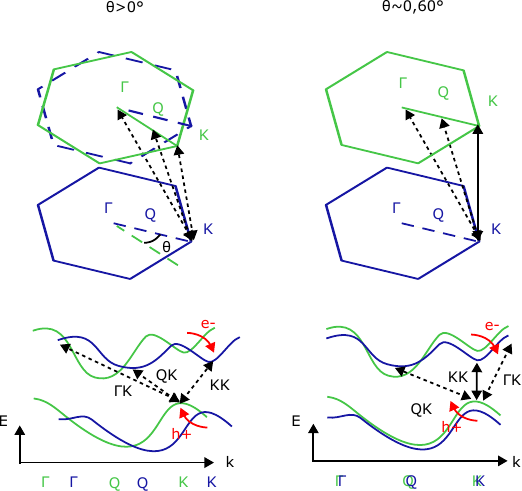}
\caption[Band structure and alignment of a TMDC heterostructure]{\textbf{Band structure and alignment of a TMDC heterostructure.} Schematic of the alignment of the Brillouin zones and band structures of two generic TMDCs. Dashes and solid lines indicate excitonic indirect and direct transitions, respectively. Red arrows indicate charge transfer.}
\label{BSScheme}
\end{figure} 

Here, we discuss for which type of heterostructures and under what conditions the sift-IX~protocol introduced in Figure~1 is applicable.\\* 

Figure~\ref{BSScheme} depicts schematically the Brillouin zone and band alignment of two generic TMDCs and the associated excitonic transitions. In this simplified picture, we neglect band hybridization and present the heterostructure band landscape as the overlap of the band structures of the constituent layers~\cite{jin2018ultrafast}. Charge transfer, i.e., the migration of opposite charges to opposite layers, is indicated with red arrows in the K valleys, although recent studies show a more complex scenario involving intervalley scattering~\cite{chan2024exciton,schmitt2022formation}. Twist angle alignment close to 0 or 60$^\circ$ makes the KK transition momentum-direct and therefore bright.\\*

The sift-IX protocol, Figure~1 of the main paper, is valid for any combination of two 2D semiconductors forming a type-II band alignment. Table I in the main text shows the intralayer and IX emission energies for the six combinations of four main TMDCs. Supplementary Table~\ref{table1} provides data for additional systems where IXs have been experimentally reported, including heterobilayers formed with non-conventional TMDCs (rhenium-based compounds and tellurides) as well as perovskites and molecules.

\begin{table}[h]
\centering
\renewcommand{\arraystretch}{1.}
\caption[Room temperature intra- and interlayer exciton energies for TMDC-based heterostructures]{\textbf{Room temperature intra- and interlayer exciton energies for TMDC-based heterostructures} The energies for the MoSe$_2$ intralayer exciton and IX in WS$_2$--MoSe$_2$ are marked with an asterisk (*) as the observed corresponding excitonic species is reported to be a hybrid of intra- and interlayer exciton. The IX in the heterostructures marked with a dagger ($^\dagger$) have only been observed below room temperature. VOPC: vanadyl-phthalocyanine. 2D perovskite: (isobutylamine)$_2$PbI$_4$}
\vspace{8pt} 
\begin{tabular}{|c|c|c|c|c|c|}
\hline
\textbf{\makecell{Combination \\ TMDC1--TMDC2}} &
\textbf{\makecell{Intralayer \\ exciton X1 \\ (eV)}} &
\textbf{\makecell{Intralayer\\ exciton X2 \\ (eV)}} &
\textbf{\makecell{IX\\ \\(eV)}} &
\textbf{\makecell{Energy\\ separation \\X2-IX (meV)}} &
\textbf{\makecell{Refs.}} \\
\hline
\textbf{MoS$_2$--MoSe$_2$} & 1.84 & 1.56 & 1.3--1.47 & 90--260 &~\cite{Alexeev2024NLa,Mouri2017Na,Zhang2018NLa} \\
\textbf{MoS$_2$--WSe$_2$} & 1.86 & 1.64 & 0.9,1.02 & 650,620 & ~\cite{Karni2019PRLa,Tan2021SAa} \\
\textbf{} & 1.85 & 1.65 & 1.6 & 50 &~\cite{Kunstmann2018NPa}  \\
\textbf{WS$_2$--MoS$_2$} & 1.98  & 1.87 & 1.45, 1.52, 1.63 & 420, 350, 240 &~\cite{Gong2014NMa, Okada2018ANa, Schottle2025APa}  \\
\textbf{WS$_2$--MoSe$_2$} & 1.98 & 1.49--1.57* & 1.49--1.57* & - &~\cite{Alexeev2019Na, Zhang2020NCa}\\
\textbf{WS$_2$--WSe$_2$} & 1.95 & 1.63 & 1.3--1.55 & 80--330 &~\cite{Jin2019NPa, Wu2021NSRa, Chen2023RAa}\\
\textbf{WSe$_2$--MoSe$_2$} & 1.66 & 1.57 & 1.33--1.38 & 190--240 &\cite{Rivera2015NCa,Nayak2017ANa, Tran2019Na, Hanbicki2018ANa,Miller2017NLa} \\
\hline
\textbf{MoS$_2$--MoTe$_2$$^\dagger$} & 1.89 & 1.15 & 0.8 & 350 &~\cite{Ju2024AMa} \\
\textbf{WSe$_2$--ReS$_2$$^\dagger$} & 1.69 & 1.55 & 1.62 & 70 &~\cite{Qin2023NLa} \\
\textbf{MoS$_2$--3-L ReS$_2$} & 1.89 & 1.67 & 1.55 & 120 &~\cite{Luo2023AAMIa} \\
\hline
\textbf{2D perovskite--WSe$_2$} & 2.3 & 1.74 & 1.58 & 160  &~\cite{yao2022layer} \\
\hline
\textbf{MoS$_2$--VOPc} & 1.85 & 1.41 & 1.54 & 310 &~\cite{Kong2022MHa} \\
\textbf{tetracene--WS$_2$$^\dagger$} & 2.3 & 2 & 1.71 & 290 &~\cite{Thompson2023Na} \\
\hline
\end{tabular}
\label{table1}
\end{table}

To ensure wide accessibility, the sift-IX protocol recommends a sequence of experimental checks and evaluation criteria that prioritizes easy implementation first and becomes increasingly demanding with subsequent steps: (i) single point spectroscopy, (ii) hyperspectral mapping, (iii) AFM, and (iv) TEPL and other methods listed in Figure 1, main text. Accordingly, all exciton energies, quenching factors, and quantitative criteria for IX compatibility based on our data or literature are room temperature values. In the following, we discuss the dependence of the protocol and the values in Table~\ref{table1} on temperature, twist angle, and material-specific strain response.  

\subsection{Quenching factors}

The sift-IX protocol (Figure~1, main text) introduces in step 1 the intralayer exciton quenching factor as a criterion to evaluate interlayer coupling in heterostructures. Based on the existing literature on interlayer coupling in TMDCs heterostructures, we propose that a threshold value of 90\% quenching in at least one of the layers defines a well-coupled heterostructure. Reported differences in quenching factors between the two constituent layers may depend on band-alignment, exciton lifetimes, and different scattering channels for electrons and holes~\cite{Volmer2023n2MaAa}. Interlayer charge transfer has also been recently reported to be sensitive to twist alignment~\cite{Zeng2025ANa,Arora2025TJoPCCa}. However, well-coupled heterostructures consistently present quenching >90\% in one of the layers. In Supplementary Table~\ref{table2}, we collect room temperature intralayer quenching factors for the six combinations of main TMDCs.

\begin{table}[h]
\centering
\renewcommand{\arraystretch}{1}
\caption[Intralayer exciton quenching factors in main TMDC bilayer combinations.]{\textbf{Intralayer exciton quenching factors in main TMDC bilayer combinations.}}
\vspace{8pt}
\begin{tabular}{|c|c|c|c|}
\hline
\textbf{\makecell{Combination \\ TMDC1--TMDC2}} &
\textbf{\makecell{Intralayer \\ exciton 1 (\%)}} &
\textbf{\makecell{Intralayer\\ exciton 2 (\%)}} &
\textbf{\makecell{Refs.}} \\
\hline
\textbf{MoS$_2$--MoSe$_2$} & 30 & 95 &~\cite{Ceballos2014ANa}\\
& 50 & 90 &~\cite{Mouri2017Na}\\
\hline
\textbf{MoS$_2$--WSe$_2$} & 90 & 97.5 &~\cite{Tan2021SAa}\\
\textbf{} & 95 & 98 &~\cite{Karni2019PRLa}\\
\hline
\textbf{WS$_2$--MoS$_2$} & 98.6 & 39 &~\cite{Chen2016NCa}\\
 & 98.3 & 73 &~\cite{Hong2014NNa}\\
\hline
\textbf{WS$_2$--MoSe$_2$} & 98.6 & 39 &~\cite{Bellus2015ANa}\\
 & 92 & 98 &~\cite{Alexeev2019Na}\\
\hline
\textbf{WS$_2$--WSe$_2$} & 100 & 96.3 &~\cite{You2022APLa}\\
 & 97.5 & 93 &~\cite{Wu2021NSRa}\\
\hline
\textbf{WSe$_2$--MoSe$_2$} & 95 & 75 &~\cite{Rivera2015NCa}\\
 & 98 & 96.7 &~\cite{Nayak2017ANa}\\
\hline
\end{tabular}
\label{table2}
\end{table}
\noindent

\subsection{Temperature dependence} The sift-IX protocol remains valid at cryogenic temperatures, considering the following points:
\begin{enumerate}
    \item Quenching factors are not expected to vary significantly with temperature because of the sub-picosecond ultrafast nature of charge transfer in TMDC heterostructures~\cite{Chen2016NCa}, as it can be observed in works reporting PL as a function of temperature~\cite{Nagler20172Ma, Lin2023NLa}
    \item Intralayer exciton energies blueshift with decreasing temperatures in a well-described manner, which is well captured by thermal models like the Varshni equation~\cite{Huang2016SRa}. This shift has little effect on the energy distances between the relevant excitons~\cite{Mouri2017Na, Ye2019Sa}.  
\end{enumerate}

\subsection{Twist angle}
In the case of momentum-direct IX, the sift-IX protocol assumes samples with the correct twist alignment. Intralayer exciton energies have been reported to change at twist angles close to 0$^\circ$ due to lattice reconstruction~\cite{lin2024moire}. The intralayer exciton energy values cited in Table~\ref{table1} are those for large twist angles.

\subsection{Material-specific strain response}
The threshold value of 200~meV below the relevant intralayer exciton corresponds to the energy range of strained intralayer excitons in realistic heterostructures. Typical strain values reported at bubbles are 1--2$\%$ of biaxial tensile strain~\cite{Khestanova2016NCa}, and exciton energy shift rates of around 100~meV/$\%$~\cite{kumar2024strain}. Since the strain response of excitons as well as adhesion forces vary across TMDCs~\cite{Carrascoso2020NRa}, the threshold value might need to be adjusted to the energy shifts observed for strained bubbles in the corresponding heterostructure if they significantly differ from the basic four TMDCs presented here. 

\newpage

\section{Supplementary Note~3: Sample statistics of the topography trap}

\begin{figure}[tp]
	\centering
	\includegraphics[width=0.98\textwidth]{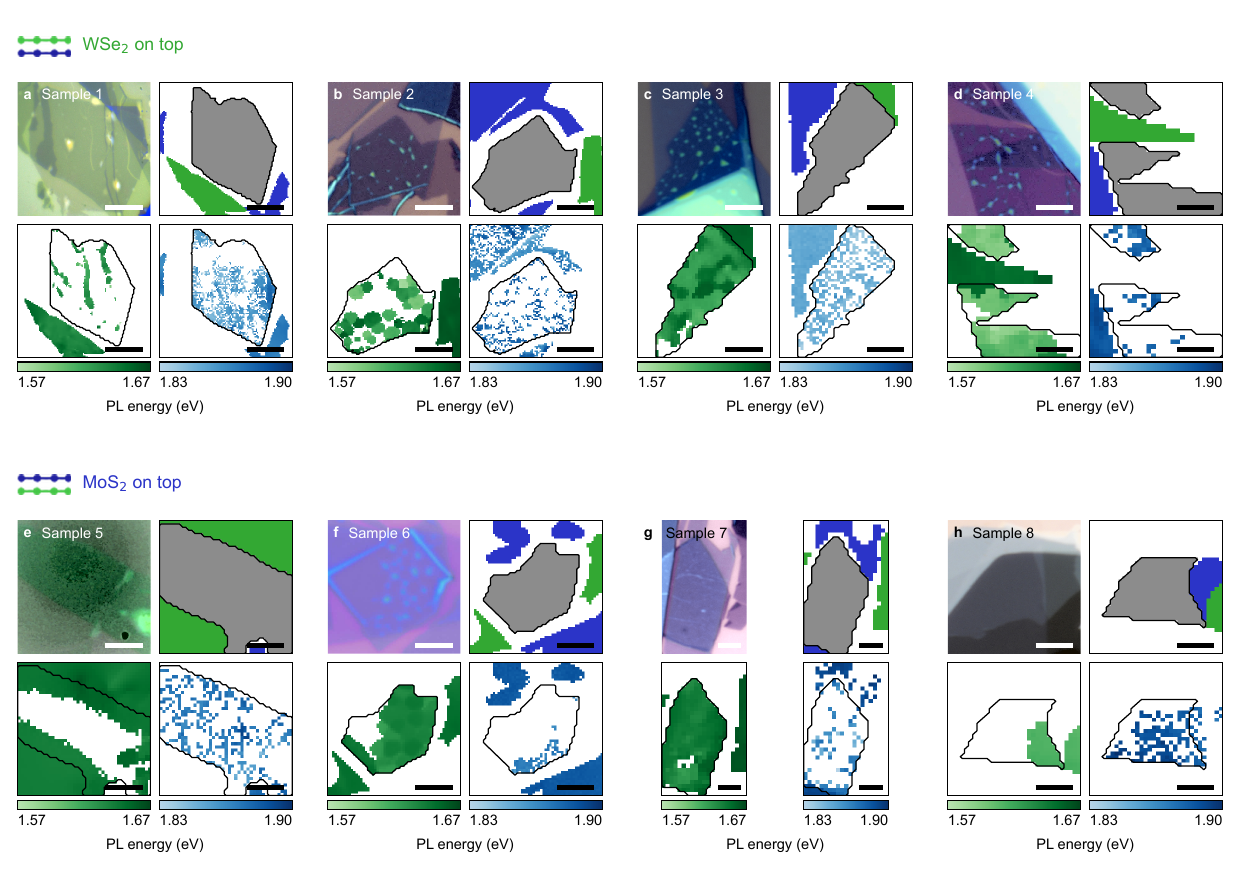}
\caption[PL energy maps of MoS$_2$--WSe$_2$ heterostructures]{\textbf{PL energy maps of MoS$_2$--WSe$_2$ heterostructures} Microscope image (top left), masks of monolayer MoS$_2$, blue, monolayer WSe$_2$, green, and overlap area, grey, (top right) and PL energy maps in  MoS$_2$ (bottom left) and WSe$_2$ (bottom right) emission ranges for eight samples, sorted by stacking order: WSe$_2$ on top (a-d) and MoS$_2$ on top (e-h). Scale bars: 11~\textmu m (a), 5~\textmu m (b), 4~\textmu m (c), 5~\textmu m (d), 3~\textmu m (e), 4~\textmu m (f), 3~\textmu m (g), 5~\textmu m (h)}. 
\label{StatsMaps}
\end{figure}

In the main text, we introduce the concept of the "topography trap": the risk of misattributing optical signals that originate from inhomogeneous strain sources, such as interface bubbles, to interlayer phenomena in TMDC-based heterostructures. Strain-inducing bubble distributions are common in real-world samples and occur across all material combinations and most fabrication techniques. A number of variables may be considered when fabricating samples. The origin and quality of the constituent monolayers (CVD, exfoliated) determines the exact energy of the intralayer excitons~\cite{liu2014strain}, the initial doping and therefore the weight of trion emission~\cite{miranda2020optical}, and the defect concentration, which can introduce additional bands in the spectrum~\cite{greben2020intrinsic}. Such effects can be measured and isolated in the reference monolayers. Other variables, such as the fabrication method (PC- or PDMS-mediated stacking), the stacking order, or the use of hBN encapsulation, have an impact on the cleanliness and homogeneity of the heterostructure interface, shaping the topography trap. In this section, we analyze more than 8000 heterostructure spectra across eight MoS$_2$--WSe$_2$ samples to provide a statistical description of the topography trap. The sift-IX protocol introduced in this work, with its empirical thresholds, and our rejection of the optical evidence for the reported $\Gamma$K-IX in MoS$_2$--WSe$_2$ are robust against different fabrication techniques, with the stacking order being the deciding parameter. \\*

\begin{figure}[htp]
	\centering
	\includegraphics[width=0.98\textwidth]{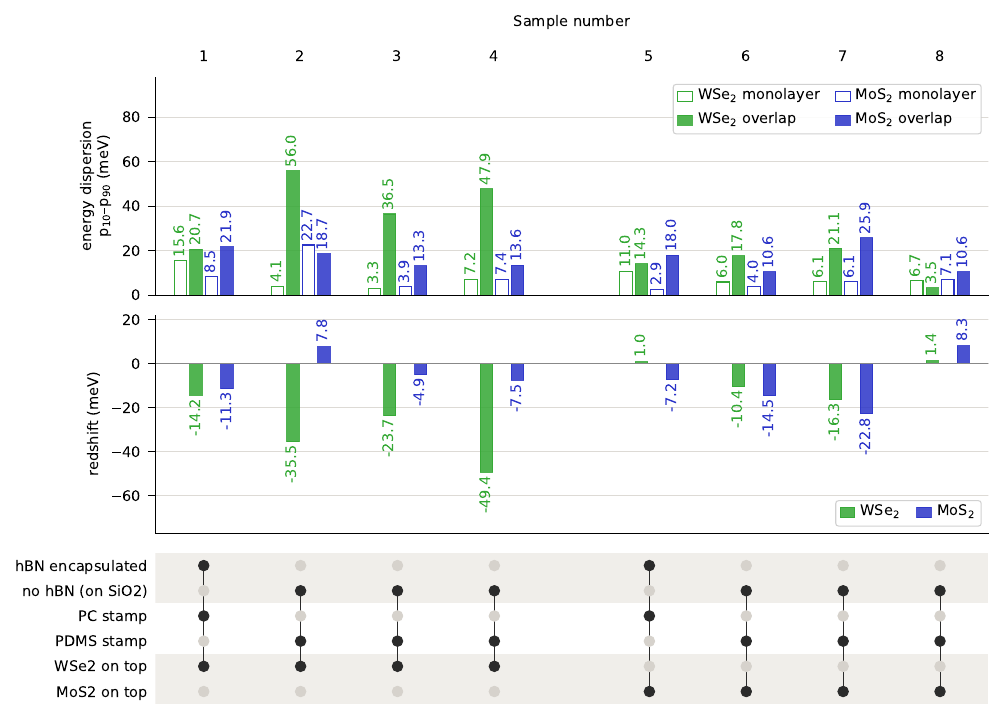}
\caption[PL energy dispersion and redshift in  MoS$_2$--WSe$_2$ heterostructures across fabrication methods]{\textbf{Energy dispersion and redshift in  MoS$_2$--WSe$_2$ heterostructures across fabrication methods} Comparison of the PL energy dispersion (energy range between the 90th and 10th percentiles of the distribution, top) and spectral shift (middle) in the overlap region compared to the monolayer (bottom), for WSe$_2$ (green) and MoS$_2$ (blue), across samples fabricated by different methods, detailed in the bottom part of the figure. Both dispersion and redshift at the overlap area are induced by bubbles in the interface, straining the top layer.}
\label{StatsEnergy}
\end{figure} 

We analyze the emission energies across eight MoS$_2$--WSe$_2$ samples for different fabrication parameters: hBN encapsulation, PC/PDMS-mediated stacking, and stacking order. As described in Figures 2 and 4 of the main text, we first inspect the integrated PL maps in the ranges 1.8--1.9\,eV (MoS$_2$) and 1.5--1.66\,eV (WSe$_2$) and compare them with microscope and AFM images to identify and mask the relevant areas: the isolated monolayers and the overlap region. In Figure~\ref{StatsMaps} we show microscope images of the studied samples and the areas identified in them, along with PL energy maps in the energy ranges corresponding to MoS$_2$ and WSe$_2$ emission. The contour of the overlap region is marked with a solid black line. Within the overlap area, we smooth the masked spectra to determine the energy of maximum counts in signals with a low signal-to-noise ratio, and we discard spurious energies coming from noisy flat spectra by checking the drop in intensity $\pm5$\,meV away from the measured peak energy. Locations with discarded spectra appear as white pixels in Figure~\ref{StatsMaps}. They are predominantly WSe$_2$ spectra for which quenching is so efficient that the spectrum is entirely flat. The energies of the surviving spectra separated by material and region were shown as histograms in Figure~6 of the main text.\\*

In Supplementary Figure~\ref{StatsEnergy}, we compare the MoS$_2$ (blue) and WSe$_2$ (green) emission energy dispersions from the monolayer and overlap regions in MoS$_2$--WSe$_2$ heterostructures for both stacking orders: WSe$_2$ (a) or MoS$_2$ (b) on top, together with their fabrication methods. We define energy dispersion as the energy range between the 90th and 10th percentiles of the distribution of energies. In general, monolayer regions show narrower dispersions of the emission energies, around 5--10\,meV, than overlap regions. This is a first indication that the energetic landscape in the overlap is more inhomogeneous than in the monolayers, due to irregularities at the interface. In the overlap, the dispersion of the WSe$_2$ energies is roughly twice as large when WSe$_2$ is on top (41~$\pm$~15~meV on average for the four samples) than when MoS$_2$ is on top (17~$\pm$~3~meV on average for the four samples), since interface irregularities strain mainly the top layer. In line with this, Figure~\ref{StatsEnergy} shows that the average WSe$_2$ energy shifts by 15--50\,meV from the monolayer to the overlap region when WSe$_2$ is on top, against 0--15\,meV when WSe$_2$ is at the bottom. These effects are milder for MoS$_2$, as its strain response is smaller and consists mainly of a drop in intensity and a growth of the indirect emission $\sim$40\,meV below the direct transition~\cite{PhysRevResearch.2.012024}, which our analysis does not capture. Samples fabricated by cleaner methods (Samples 1 and 5) follow the same trends, but more weakly, as the clean PC stamp leaves fewer bubbles and the hBN reduces the inhomogeneity.\\*

The results in this section show the robustness of our description of the topography trap against sample fabrication methods. Across the samples investigated here, stacking order is the dominant correlate of the WSe$_2$ energy shift and dispersion --- the same qualitative trend is observed across the fabrication approaches represented --- as interfacial bubbles strain mainly the top layer. Furthermore, the effects ascribed to strain do not depend on the substrate; that is, strain dominates over any other possible interlayer coupling effect with the bottom layer, as we show in Supplementary Note~6. The rich inhomogeneous strain landscape that underlies the topography trap is therefore generic to manually stacked TMDC heterostructures. The sift-IX protocol introduced in Figure 1 has broad applicability, not only for telling interlayer excitons from artifacts but also for the full characterization of real strain-sensitive excitons.

%\pagebreak
\newpage

\section{Supplementary Note~4: IX screening methods beyond the sift-IX protocol}

In Figure~1 of the main text, the sift-IX protocol suggests three follow-up experimental checks when the analyzed PL feature is found to be a topographical artifact: nanoscale spectroscopy, AFM interface cleaning, and inverting the stacking order. In Supplementary Figure~\ref{Squeegee}, we demonstrate contact-mode AFM cleaning. In Supplementary Figure~\ref{StackKK}, we demonstrate the stacking order check with a positive case, the KK-IX of MoS$_2$-WSe$_2$. In Supplementary Figure~\ref{MoS2TEPL} we perform TEPL on a MoS$_2$-WSe$_2$ heterostructure.

\subsection{Contact-mode AFM interface cleaning}

As discussed in the main text, ambient atmosphere and polymer-related contaminants get trapped at the heterostructure interface during dry-transfer fabrication. These residues physically separate the layers and impede interlayer coupling. A common strategy to achieve flat, well-coupled areas is thermal annealing. Annealing increases the mobility of trapped residues and triggers a self-cleaning mechanism in which residues coalesce in pockets, leaving pristine areas behind. However, this process offers little control over the size and distribution of the resulting well-coupled areas.

\begin{figure}[tp]
	\centering
	\includegraphics{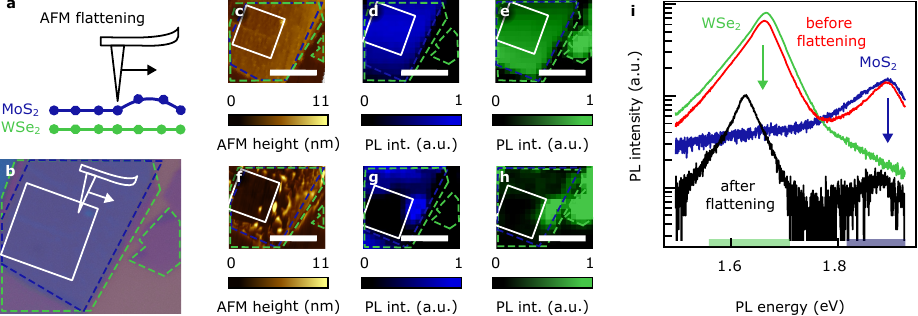}
\caption[Contact-mode AFM interface cleaning]{\textbf{Contact-mode AFM interface cleaning.} We push aside interfacial contaminants at the interface of a MoS$_2$/WSe$_2$ heterostructure with an AFM tip in contact mode. This results in an area of clean interface with efficient interlayer coupling, demonstrated by 90\% intralayer exciton quenching in the flattened area with respect to the isolated layers. \textbf{a} Schematic depicting AFM flattening. \textbf{b} Microscope image of the MoS$_2$/WSe$_2$ heterostructure before cleaning. \textbf{c} and \textbf{f} AFM images of the sample before and after cleaning of the area marked with a white square. \textbf{d,e} and \textbf{g,h} PL intensity map of intralayer MoS$_2$ (d,g) and WSe$_2$ (e,h) excitons in the sample before and after cleaning, respectively. Integration energies for the corresponding maps are indicated in \textbf{i} Representative PL spectra (log scale) of the sample. Scale bars: 2~\textmu m.}
\label{Squeegee}
\end{figure} 

Alternatively, pristine interfaces can be achieved by actively manipulating interlayer residues between layers using an AFM tip in contact mode. The concept is illustrated in Figure~\ref{Squeegee}a. An AFM tip pushes pockets of contaminants aside to engineer large, atomically flat areas. We demonstrate this technique on a MoS$_2$/WSe$_2$ heterostructure shown in Figure~\ref{Squeegee}b.

We start by characterizing the as-transferred heterostructure in contact-mode AFM with a low applied force of 2~nN (which does not affect interfacial contaminants) in Figure~\ref{Squeegee}c and PL hyperspectral mapping in Figures~\ref{Squeegee}d,e. MoS$_2$ and WSe$_2$ intralayer exciton PL intensities in Figures~\ref{Squeegee}d,e are integrated in the ranges indicated in Figure~\ref{Squeegee}i. The initial height step between the heterostructure and the substrate is 7 nm, more than 4 times larger than the expected 1.5 nm for a pristine bilayer. This strongly indicates the presence of a layer of interfacial residues. Consistently, the PL of the overlap region in Figure~\ref{Squeegee}i (red) is the sum of the unquenched intralayer excitons from the two non-interacting, uncoupled layers.\\

In Figure~\ref{Squeegee}f, we image the sample after clearing the residues from a square region using contact AFM, see Methods in Supplementary Note~1. The height of the heterostructure in the squared area is now around 1.6~nm, close to the bilayer limit. Displaced contaminants accumulate in pockets on the edge of the scanned area. PL spectra from the cleaned area, Figure~\ref{Squeegee}i(black), and maps of MoS$_2$ and WSe$_2$ intralayer exciton PL, Figures~\ref{Squeegee}g,e, reveal that AFM cleaning causes an order of magnitude PL intensity decrease. This PL quenching arises from charge transfer due to efficient interlayer coupling,  which was absent before the cleaning step. Consistent with Figures~2 and 4 of the main paper, as well as all other findings of this work, we find no evidence of $\Gamma$K-IX.\\*
\newpage

\subsection{Inverted stacking order} 

\begin{figure}[tp]
	\centering
	\includegraphics[width=0.98\textwidth]{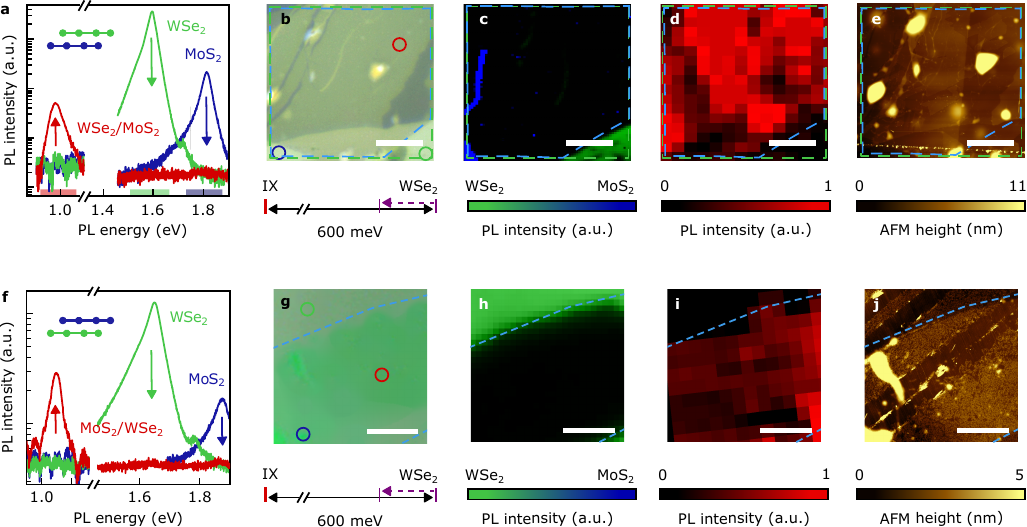}
\caption[Stacking order check of the KK-IX in MoS$_2$--WSe$_2$.]{\textbf{Stacking order check of the KK-IX in MoS$_2$--WSe$_2$.} We fabricate two MoS$_2$-WSe$_2$ heterostructures with opposite stacking orders to test the robustness of the KK-IX assignment. \textbf{a, f} Representative log scale PL spectra collected at the points marked with the same color in \textbf{b, g} Microscope images of MoS$_2$-WSe$_2$  heterostructures with MoS$_2$ (b) and WSe$_2$ (g) on top and corresponding schematics of the energy distance between IX and the nearest intralayer exciton. \textbf{c, h} PL intensity maps of intralayer excitons. \textbf{d,i} PL intensity maps of IX. \textbf{e,j} AFM topography. Scale bars: a--e: 10~\textmu m, f--j: 5~\textmu m.}
\label{StackKK}
\end{figure} 

An additional requirement for IX-compatible PL signatures is to be present in heterostructures independently of the stacking order. We demonstrated this experimental check in Figure~3 of the main text, showing that the $\Gamma$K-IX-compatible emission in WS$_2$--WSe$_2$ appears for both stacking orders, while in Figure~4 of the main text we show that the $\Gamma$K-IX in MoS$_2$--WSe$_2$ is not robust against stacking order inversion. Here, we add another IX that passes the test, the KK-IX in MoS$_2$--WSe$_2$.\\*

\begin{figure}[tp]
	\centering
	\includegraphics{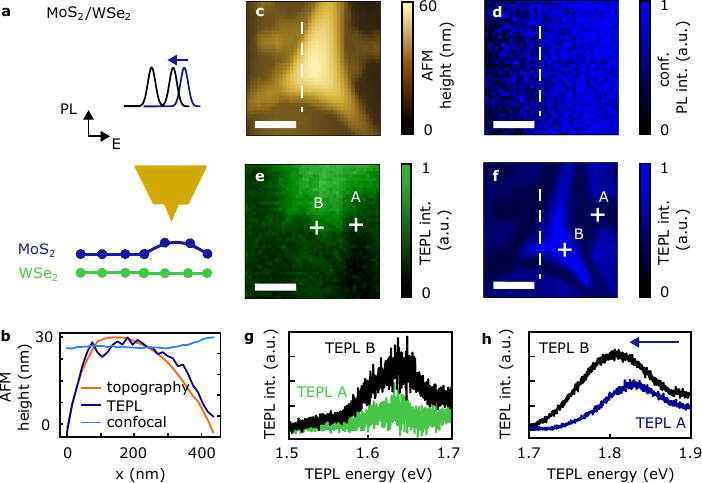}
\caption[TEPL of a MoS$_2$/WSe$_2$ heterostructure]{\textbf{TEPL of a MoS$_2$/WSe$_2$ heterostructure} In Figure 3 of the main text, we use TEPL to resolve the emission of a bubble in a WSe$_2$/MoS$_2$ heterostructure. Here, we perform TEPL on a sample with the opposite stacking order, MoS$_2$/WSe$_2$, to further investigate the effect of interlayer bubbles on the heterostructure's top-layer emission. \textbf{a} Schematic of the TEPL experiment and expected PL spectra from the heterostructure (black) compared to that of unstrained isolated MoS$_2$ (blue). \textbf{b} Correlated AFM topography, confocal PL, and TEPL intensity linescans across a bubble, marked with white dashed lines in (c), (d), and (f), demonstrating near-field spatial resolution. \textbf{c} AFM topography of the interlayer bubble. \textbf{d, f} Far-field and near-field PL intensity maps integrated over the MoS$_2$ exciton. \textbf{e} Near-field PL intensity map of the WSe$_2$ exciton. \textbf{g,h} Representative TEPL spectra corresponding to the locations marked in (e) and (f), respectively. On the bubble, the top-layer emission (MoS$_2$) is downshifted due to strain, in contrast to the flat bottom-layer emission (WSe$_2$). Scale bars: 180~nm.}
\label{MoS2TEPL}
\end{figure}

Two MoS$_2$-WSe$_2$ heterostructures with opposite stacking orders are shown in Figure~\ref{StackKK}: the sample in Figures~\ref{StackKK}a-e has WSe$_2$ on top; the sample in Figures~\ref{StackKK}f-j has MoS$_2$ on top. The sample in Figures~\ref{StackKK}a-e has already been discussed in Figure~2 of the main text. Both samples are hBN-encapsulated and aligned to a twist angle close to 0$^\circ$ to observe KK-IX. For both stacking orders, we observe an emerging KK-IX around 1~eV (Figures~\ref{StackKK}a and f), present everywhere in the overlap area (Figures~\ref{StackKK}d and i), and fully quenched intralayer excitons (Figures~\ref{StackKK}c and h). The topography irregularities in Figures~\ref{StackKK}e and j do not cancel intralayer exciton quenching or the presence of IX and are therefore likely to be contaminants trapped between the hBN and the heterostructure. The KK-IX in MoS$_2$-WSe$_2$ is robust against stacking order inversion. We could not find any emission below the intralayer WSe$_2$ exciton to be attributed to the $\Gamma$K-IX in either sample.\\*

Finally, in Figure~5 of the main text, we use TEPL to show how a bubble of contaminants between WSe$_2$ and MoS$_2$ in a WSe$_2$/MoS$_2$ heterostructure strains the top layer, producing an unquenched, redshifted WSe$_2$ exciton signal. In Figure~\ref{MoS2TEPL}, we perform TEPL on a bubble in a MoS$_2$/WSe$_2$ heterostructure, to show how, in this case, when the MoS$_2$ is on top, the signal from the bubble is an unquenched, redshifted MoS$_2$ exciton.

\newpage

\section{Supplementary Note~5: Twist-angle-dependence test of MoS$_2$--WSe$_2$ and WS$_2$--WSe$_2$ heterostructures}

\begin{figure}[tp]
	\centering
	\includegraphics{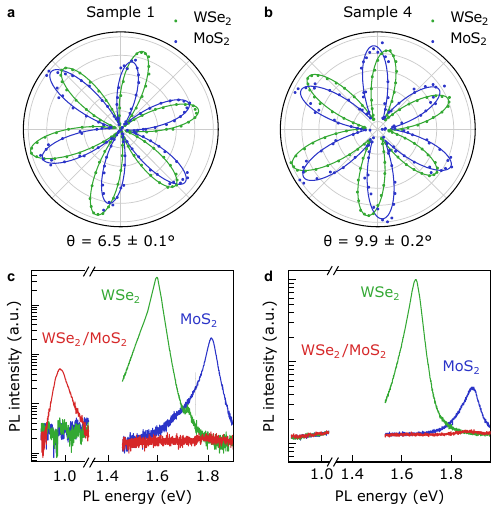}
\caption[Twist-angle-dependence test of MoS$_2$--WSe$_2$ heterostructures]{\textbf{Twist-angle-dependence test of MoS$_2$--WSe$_2$ heterostructures} SHG measurement of the twist angle for two WSe$_2$/MoS$_2$ heterostructures. \textbf{c,d} Representative PL spectra from the isolated monolayers (WSe$_2$, green and MoS$_2$, blue) and overlap area (red) for the two heterostructures measured in (a,b). The sample with a smaller twist angle shows a KK-IX at 1~eV and no $\Gamma$K-IX, while the sample with a larger twist angle shows no IX at all.}
\label{SHG}
\end{figure} 

\begin{figure}[tp]
	\centering
	\includegraphics{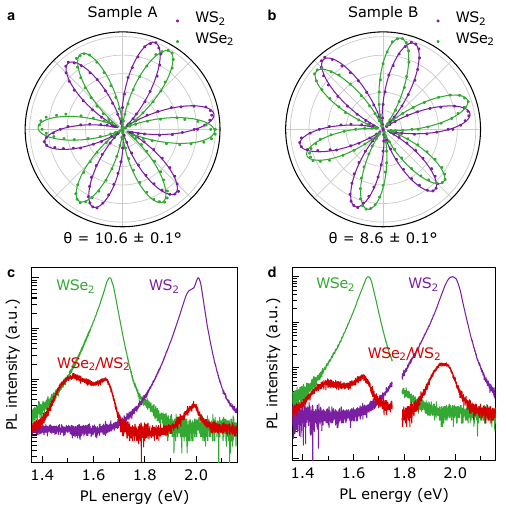}
\caption[Twist-angle-dependence test of WS$_2$--WSe$_2$ heterostructures]{\textbf{Twist-angle-dependence test of WS$_2$--WSe$_2$ heterostructures} \textbf{a,b} SHG measurement of the twist angle for two WSe$_2$/WS$_2$ heterostructures. \textbf{c,d} Representative PL spectra from the isolated monolayers (WSe$_2$, green and WS$_2$, purple) and overlap area (red) for the two heterostructures measured in (a,b). Both samples, with different twist angles, show the spectral feature below the WSe$_2$ intralayer exciton assigned to a $\Gamma$K-IX. }
\label{SHG2}
\end{figure}

In the main text, we study two interlayer excitons in MoS$_2$--WSe$_2$ heterostructures: the momentum-direct, twist-angle-dependent KK-IX and the momentum-indirect, twist-angle-independent $\Gamma$K-IX. The KK-IX requires the K valleys of the constituent layers to be aligned, see Figure~\ref{BSScheme}, so its intensity is maximum at $\theta = 0, 60^\circ$ and decays rapidly with twist angle~\cite{Rivera2018NNa}. The $\Gamma$K-IX, in contrast, has been reported to emit with an intensity that is constant in twist angle~\cite{Kunstmann2018NPa}. The two excitons therefore respond differently to differences in twist angle. Here, we determine with SHG the twist angle of two MoS$_2$--WSe$_2$ heterostructures and compare their interlayer exciton spectra. We compare the result for the reported $\Gamma$K-IX in MoS$_2$--WSe$_2$ to that of WS$_2$--WSe$_2$. See Supplementary Note~1 for details on the SHG measurements.\\*

In Figure~\ref{SHG}a we measure the twist angle of the heterostructure of Figure~2 of the main text. This sample was stacked nominally aligned; SHG gives a twist angle of $\theta = 6.5^\circ$. This lies within the range in which direct IXs have been reported~\cite{Karni2019PRLa} and also within the range in which an indirect IX is predicted to emit~\cite{Kunstmann2018NPa}. However, in this sample we find clear evidence of the KK-IX but none of the $\Gamma$K-IX, Figure~\ref{SHG}c.\\*

To test this further, we fabricate a second heterostructure, this time deliberately misaligned, and measure both excitons again. SHG gives $\theta = 9.9^\circ$ for this sample, Figure~\ref{SHG}c. As shown in Figure~\ref{SHG}~d, the KK-IX is now gone, as expected from its twist-angle dependence~\cite{Nayak2017ANa}, while the $\Gamma$K-IX remains absent even though it should be present at all twist angles.\\*

For comparison, we fabricate two WS$_2$--WSe$_2$ heterostructures with different stacking angles, see Figure~\ref{SHG2}. We measure their twist angle via SHG and find in Figure~\ref{SHG2}a $\theta = 10.6^\circ$ for sample A (also shown in Figure~3a-f of the main text), and $\theta = 8.6^\circ$ for sample B, Figure~\ref{SHG2}b. We observe in both samples an emerging feature below the WSe$_2$ at 1.52~eV (Figure~\ref{SHG2}c) and 1.50~eV (Figure~\ref{SHG2}d) that is compatible with the reported $\Gamma$K-IX assignment. The relative energies and twist angles agree with those reported in ref.\cite{Wu2021NSRa}.

\newpage
\section{Supplementary Note~6: Interface bubbles between WSe$_2$ and a gold substrate}

\begin{figure}[tp]
	\centering
	\includegraphics{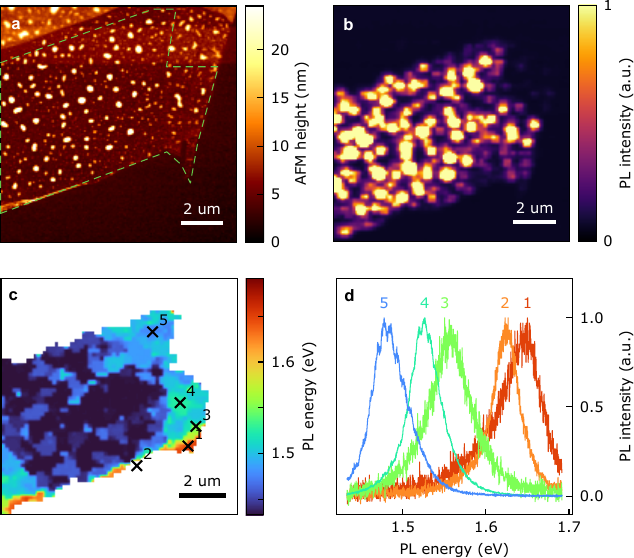}
\caption[Interface bubbles between WSe$_2$ and a gold substrate]{\textbf{Interface bubbles between WSe$_2$ and a gold substrate} Contaminant-filled bubbles between a WSe$_2$ monolayer and a gold substrate show the same redshifted WSe$_2$ PL around bubbles as MoS$_2$--WSe$_2$ heterostructures. \textbf{a} AFM map of a WSe$_2$ flake on a gold substrate featuring interface bubbles. The monolayer WSe$_2$ is contoured with a green dashed line. \textbf{b,c} PL intensity and energy map showing PL emission in the vicinity of bubbles in different energy ranges depending on local strain. \textbf{d} Normalized representative spectra from the locations marked in (c). The energy range covered by the strained WSe$_2$ bubbles coincides with the strain range in MoS$_2$--WSe$_2$ heterostructures.}
\label{AuBubble}
\end{figure} 

In Figures 3 and 4 of the main text, we explain how bubbles filled with polymer residues from fabrication and trapped at the interface of heterobilayers strain the top layer, redshifting its emission. This redshifted PL from the strained top layer in MoS$_2$--WSe$_2$ heterostructures has often been misattributed in the literature to interlayer excitons. To further show that the PL from the vicinity of bubbles is due to strain and not to interactions with the bottom MoS$_2$ layer, we fabricate samples in which such bubbles are trapped between WSe$_2$ and a gold substrate.  WSe$_2$ PL emission is suppressed on the metallic substrate due to charge transfer~\cite{rodriguez2022PRB}, but remains bright on bubbles where the monolayer detaches. The bubbles strain the WSe$_2$ and redshift its emission, producing optical signatures similar to those in MoS$_2$--WSe$_2$ but here without any MoS$_2$ layer.\\*

We prepare the samples by PDMS stamping of exfoliated WSe$_2$ on Au. We deliberately use PDMS to introduce the residues that, after annealing, coalesce into bubbles comparable to those in the MoS$_2$--WSe$_2$ samples. Figure \ref{AuBubble}a shows an AFM scan of a WSe$_2$/Au sample, with the WSe$_2$ monolayer contour marked by a green dashed line. We identify the protuberances in the monolayer region as bubbles of contaminants trapped between the WSe$_2$ and the substrate by comparing with the WSe$_2$ PL intensity map, Figure \ref{AuBubble}b. WSe$_2$ PL intensity increases on these bubbles as the monolayer detaches and charge transfer to the metallic substrate is suppressed. In Figure \ref{AuBubble}d we plot representative spectra taken at the positions marked with crosses in the PL energy map, Figure \ref{AuBubble}c. Interfacial bubbles strain the WSe$_2$ by different amounts, redshifting its emission over the same 1.5-1.65\,eV range reported in the main text for MoS$_2$--WSe$_2$ heterostructures.\\*

\newpage

\section{Supplementary Note~7: Summary of arguments against the reported $\Gamma$K-IX assignment in MoS$_2$--WSe$_2$}
From the application of the sift-IX protocol in Figure~3, it follows that the assignment of the emission below the pristine WSe$_2$ intralayer exciton to an IX in MoS$_2$--WSe$_2$ heterostructures is inconsistent with interlayer emission. Here, we summarize the arguments against the reported existence of optical signatures for $\Gamma$K-IX in MoS$_2$--WSe$_2$:

\begin{enumerate}
\item \textbf{Reduced intralayer quenching.} MoS$_2$ intralayer emission increases whenever a $\Gamma$K-IX--like feature emerges, both in our data (Figure~4d) and whenever this is shown in~\cite{Kunstmann2018NPa} and~\cite{Fang2014PotNAoSa,Chiu2014ANa, Nagler2019pssba, Cho2021NLa, Ren2022APLa, Khestanova2023APa}. Reduced intralayer quenching indicates reduced charge transfer and, therefore, the dissociation of interlayer excitonic species.
    
\item \textbf{Spatial correlation with topography.} The $\Gamma$K-IX--like feature correlates spatially with topographical irregularities such as interface bubbles that hinder interlayer coupling, compare for instance Figures~4d and e or Figures~5b and c.

Additionally, TEPL measurements (Figure~5e, f) localize the PL below the energy of the unstrained WSe$_2$ exciton exclusively to WSe$_2$ bubbles while PL from the surrounding, well-coupled areas is completely quenched.

\item \textbf{Samples with KK-IX show no $\Gamma$K-IX.} The $\Gamma$K-IX--like feature is absent in samples where the observation of KK-IX proves good interlayer coupling, see Figures~2f, ~\ref{StackKK}, ~\ref{SHG}. The theoretical framework of the $\Gamma$K IX does not support a scenario in which the exciton is absent for $\theta \sim 0^\circ, 60^\circ$ but present for all other angles.

\item \textbf{Stacking order.} Tellingly, in the vast majority of $\Gamma$K-IX reports, the WSe$_2$ layer is on top. Because the contaminant bubbles in the interface strain mainly the top layer, only when WSe$_2$ is on top, the bubbles strain it enough to produce the spectral features that can be mistaken with IX. In Figures~4 and 6 of the main text, we show PL data for devices with opposite stacking order. Indeed, when WSe$_2$ is on top, the bubbles in the interface redshift its intralayer exciton, while it remains largely unchanged when WSe$_2$ is on the bottom. 

\item \textbf{Interface cleanliness.} The $\Gamma$K-IX-like feature is absent in samples where interface cleanliness is directly realized via contact-mode AFM nanoflattening, see Figure~\ref{Squeegee}. 

\item \textbf{Energy dispersion consistent with strain.} $\Gamma$K-IX is reported at widely differing energies between 1.5~eV and 1.6~eV~\cite{Kunstmann2018NPa}, consistently with a random distribution of bubbles and strain, as we compare in Figure~5f.

\item \textbf{ARPES.} Finally, angle-resolved photoemission spectroscopy on MoS$_2$--WSe$_2$ heterostructures has found no evidence of the $\Gamma$K-IX~\cite{Karni2022Na, schmitt2022formation}.

\end{enumerate}

%\clearpage
% Don't count these!
%TC:ignore
%\quickwordcount{supplement_resubmission}
%\quickcharcount{MoS2metals}
%\detailtexcount{supplement_resubmission}

\clearpage

\newpage
\bibliographystyle{achemso}
\bibliography{acs-control,20261020TopographyTrap_fixed}